\documentclass[fleqn,usenatbib,useAMS]{mnras}

\usepackage{graphicx}	
\usepackage{amsmath}	
\usepackage{multicol}        
\usepackage{bm}		
\usepackage{pdflscape}	
\usepackage{xcolor}
\usepackage{soul}
\usepackage{booktabs}
\usepackage[capitalise]{cleveref}
\usepackage{tikz}
\usepackage{float}

\usetikzlibrary{shapes.geometric, arrows, calc, positioning, fit}
\tikzset{every node/.style={inner sep=10pt,minimum height=1cm}}
\tikzstyle{startstop} = [rectangle, rounded corners, minimum width=1cm, minimum height=1cm, text centered, draw=black, fill=blue!50]
\tikzstyle{io} = [trapezium, trapezium left angle=70, trapezium right angle=110, minimum width=1cm, text width=5cm, minimum height=1cm, text centered, draw=black, fill=blue!30]
\tikzstyle{process} = [rectangle, minimum width=1cm, minimum height=1cm, text centered, text width=6cm, draw=black, fill=gray!20]
\tikzstyle{middle} = [trapezium, trapezium left angle=80, trapezium right angle=100, minimum width=0cm, text width=6cm, minimum height=1cm, text centered, draw=black, fill=green!10]
\tikzstyle{middle2} = [trapezium, trapezium left angle=80, trapezium right angle=100, minimum width=0cm, text width=6cm, minimum height=1cm, text centered, draw=black, fill=green!10]
\tikzstyle{subio} = [rectangle, minimum width=1cm, minimum height=1cm, text centered, draw=black, fill=blue!30]
\tikzstyle{arrow} = [thick,->,>=stealth]
\tikzstyle{branchframe} = [draw=black!55, dashed, rounded corners, inner sep=6pt]

\usepackage[T1]{fontenc}
\usepackage{ae,aecompl}

\usepackage{newtxtext,newtxmath}

\title[Field-level cosmology with multifidelity SBI]{Field-level weak lensing cosmology with $60$ simulations using multifidelity simulation-based inference}

\author[A. A. Saoulis et al.]{Alex A. Saoulis,\thanks{Contact e-mail: \href{mailto:a.saoulis@ucl.ac.uk}{a.saoulis@ucl.ac.uk}}$^{1,2}$ 
Kiyam Lin,$^1$
Niall Jeffrey,$^{3,1}$ 
Maximilian von Wietersheim-Kramsta,$^4$ 
Davide Piras,$^{5,6}$ 
\newauthor
Alessio Spurio Mancini,$^{7}$ 
Ana M. G. Ferreira,$^2$
Benjamin Joachimi$^1$
\\
$^1$ Department of Physics \& Astronomy, University College London, Gower Street, London, WC1E 6BT, United Kingdom \\
$^2$ Department of Earth Sciences, University College London, 5 Gower Place, London, WC1E 6BS, United Kingdom  \\
$^3$ Department of Physics \& King's Institute for Artificial Intelligence, King’s College London, Strand, London WC2R
2LS, United Kingdom \\
$^4$ Institute for Computational Cosmology (ICC) \& the Centre for Extragalactic Astronomy (CEA), Durham University, Durham, United Kingdom  \\
$^5$ Département de Physique Théorique, Université de Genève, 24 quai Ernest Ansermet, 1211 Genève 4, Switzerland\\
$^6$ ETH Zurich, Institute for Particle Physics and Astrophysics, Wolfgang-Pauli-Strasse 27, 8093 Zurich, Switzerland\\
$^7$ Department of Physics, Royal Holloway, University of London, Egham Hill, Egham, TW20 0EX, United Kingdom
  }

\date{}

\pubyear{{\the\year{}}}

\begin{document}
\label{firstpage}
\pagerange{\pageref{firstpage}--\pageref{lastpage}}
\maketitle

\begin{abstract}
We perform a realistic KiDS-Legacy mock analysis with field-level neural compression and simulation-based inference using just 60 $N$-body simulations. The weak lensing shear field encodes substantially more cosmological information than standard two-point summary statistics such as the power spectrum. Field-level inference can fully exploit this information, but physical realism at the field-level requires very high-fidelity simulations. This poses a major challenge for simulation-based inference (SBI): accurate empirical density modelling and deep-learning-based neural compression require tens of thousands of training samples, but achieving physical realism at the field level makes each simulation extremely costly. We demonstrate that multifidelity SBI can alleviate this tension by substantially reducing the number of high-fidelity simulations needed for accurate cosmological inference. We pre-train neural inference models on realistic KiDS-Legacy-like shear mocks using fast log-normal \texttt{GLASS} simulations and fine-tune them on a small set of high-fidelity $N$-body simulations. We show that $60$ high-fidelity simulations are sufficient to obtain informative and well-calibrated cosmological posteriors, enabling at least an order-of-magnitude reduction in simulation cost for accurate field-level inference in a realistic setting.

\end{abstract}

\begin{keywords}
cosmology: cosmological parameters, large-scale structure of Universe, dark matter -- methods: statistical - software: machine learning 
\end{keywords}



\section{Introduction}

Stage-III cosmological surveys, such as the Kilo-Degree Survey \citep[KiDS;][]{kuijken2015gravitational,wright2024fifth}, the Dark Energy Survey \citep[DES;][]{abbott2018dark}, and the Hyper Suprime-Cam \citep[HSC;][]{aihara2018hyper, hikage2019cosmology} survey, have produced high resolution mappings of galaxies and their properties that have enabled increasingly precise cosmological constraints \citep{abdalla2022cosmology, di2025cosmoverse}. Weak lensing measurements of these galaxies provide a relatively clean probe of cosmic structure over large time and length-scales. This has enabled analysis of key cosmological questions, providing independent constraints on the composition of the Universe and the behaviour of gravity and dark energy \citep{wright2025kids, novaes2025cosmology, cheng2025cosmological, jeffrey2025dark, prat2026dark, abbott2025dark, abbott2026dark, stolzner2026kids}. 

We identify two significant challenges in taking full advantage of the wealth of cosmological information in these data. The first is extracting information beyond the traditional two-point statistics, such as correlation functions or the power spectrum, as we probe smaller and more non-linear scales \citep{takada2003three,dietrich2010cosmology,kratochvil2012probing,harnois2021cosmic}. The second challenge, directly related to the first, is ensuring accurate physical modelling of the evolution of cosmic structure, particularly at these small scales \citep{schneider2020baryonic,arico2023y3, elbers2025flamingo}. Addressing these challenges will only increase in importance for Stage-IV surveys that will produce even higher quality galaxy catalogues \citep[e.g. \textit{Euclid}, see][]{mellier2024euclid,martinelli2021euclid, ajani2023euclid, castro2024euclid,navarro2026euclid}. 

Regarding the first challenge, higher-order statistics have now seen widespread adoption and success.  Including non-Gaussian information can significantly improve constraints, for instance on the matter density parameter $\Omega_\mathrm{m}$ \citep[e.g.,][]{novaes2025cosmology, jeffrey2025dark}.  One can use statistics (such as peak counts) of the convergence fields \citep{gatti2022dark, giblin2023enhancing} or of aperture mass $M_\mathrm{ap}$ maps \citep{harnois2021cosmic,harnois2024kids, marques2024cosmology}. Three-point statistics are another popular approach; for instance, the third moment of the aperture mass, $\langle M_\mathrm{ap}^3 \rangle$, is a convenient probe of the three point function \citep{secco2022dark,zurcher2022dark,burger2024kids,gomes2025dark}. Prior work has also demonstrated the value of statistics such as persistent homology \citep{heydenreich2022persistent,prat2026dark} and the wavelet scattering transform \citep{regaldo2024galaxy,cheng2025cosmological}. This broad suite of powerful, complementary, yet often insufficient \citep{novaes2025cosmology, sui2026evaluate} summary statistics motivates a more generic approach to constructing optimal summaries.

To this end, data-driven, machine learning (ML)-based approaches for extracting information from cosmological measurements are promising. Deep neural networks can be trained to learn optimal summaries from more complete cosmological observations. These can be applied to pixelised, map-level representations of the lensing field \citep{gupta2018non, ribli2019weak, fluri2019cosmological,  fluri2022full,  matilla2020interpreting, makinen2021lossless, jeffrey2021likelihood,lu2023cosmological, lanzieri2024optimal, lemos2024field} or directly to catalogue-like, point-cloud products \citep{de2023robust, lehman2024learning, thomsen2024des}. Designing effective data representations and neural network architectures remains a very active area of research \citep{villanueva2022learning,dai2024multiscale,jeffrey2025dark, thomsen2024des}. Nonetheless, recent applications of ML-based compression have produced the tightest weak lensing constraints on the matter composition and dark energy sector to-date \citep{lu2023cosmological, jeffrey2025dark, prat2026dark}. 

We turn to the second challenge. As information beyond-two-point statistics is included in inference, utmost care must be taken in ensuring accurate physical modelling (and, by extension, statistical modelling through the likelihood). On small, non-linear scales, the dynamics of structure formation are governed by complex gravitational and baryonic processes that are difficult to model analytically \citep{schneider2020baryonic, arico2023y3, elbers2025flamingo}. As a result, accurate forward modelling heavily relies on large-volume $N$-body or hydrodynamical simulations \citep{weinberger2016simulating, villaescusa2020quijote, maksimova2021abacussummit, kacprzak2023cosmogridv1, schaye2023flamingo, jeffrey2025dark}, sometimes augmented with targeted high-resolution techniques such as zoom-in or in-painting methods \citep{hahn2022rm, hahn2024cosmological, nadler2025cozmic, chatterjee2025cosmology, schneider2025baryonification}. While these simulations provide high-fidelity predictions, their computational cost typically precludes their direct use within inference, and they are instead used to emulate or calibrate simplified likelihoods for a given summary statistic \citep{mead2016accurate,bacco,euclid2021euclid}.

A natural framework for addressing both the extraction of information from complex data and the incorporation of realistic physical modelling is simulation-based inference (SBI; also known as likelihood-free or implicit-likelihood inference). Here, we use the term SBI to refer specifically to methods that replace an explicit likelihood with a data-driven neural density estimator (NDE) trained on forward simulations to learn the relationship between cosmological parameters and observables (e.g., \citealt{papamakarios2016fast, alsing2019fast, cranmer2020frontier}, as opposed to explicit likelihood emulation approaches such as \citealt{giblin2023enhancing,marques2024cosmology}). This enables inference with high-dimensional, non-Gaussian summaries without restrictive assumptions on the likelihood, while systematically incorporating observational effects and non-linear physics through the simulations themselves. However, the accuracy of SBI is fundamentally limited by the number and fidelity of simulations available for training, making direct application with state-of-the-art simulations computationally prohibitive \citep{park2025dimensionality, bairagi2025many, saoulis2025transfer, krouglova2025multifidelity}.

Multifidelity methods have emerged as a means to directly utilise high-fidelity simulations during inference \citep{lee2024zooming,jia2024simulation, jia2024cosmological, krouglova2025multifidelity, hikida2025multilevel, saoulis2025transfer}. These leverage lower fidelity simulations to learn either or both of i) the information content in the cosmological observables, for instance through neural compression and ii) the probabilistic structure of the observables, through for instance the structure of the likelihood or posterior. These low-fidelity models can then be rapidly adapted to the higher fidelity problem using a much smaller number of high-fidelity examples. These approaches rely on ML and are therefore well-suited to SBI techniques.

 \begin{figure}
    \includegraphics[width=\columnwidth]{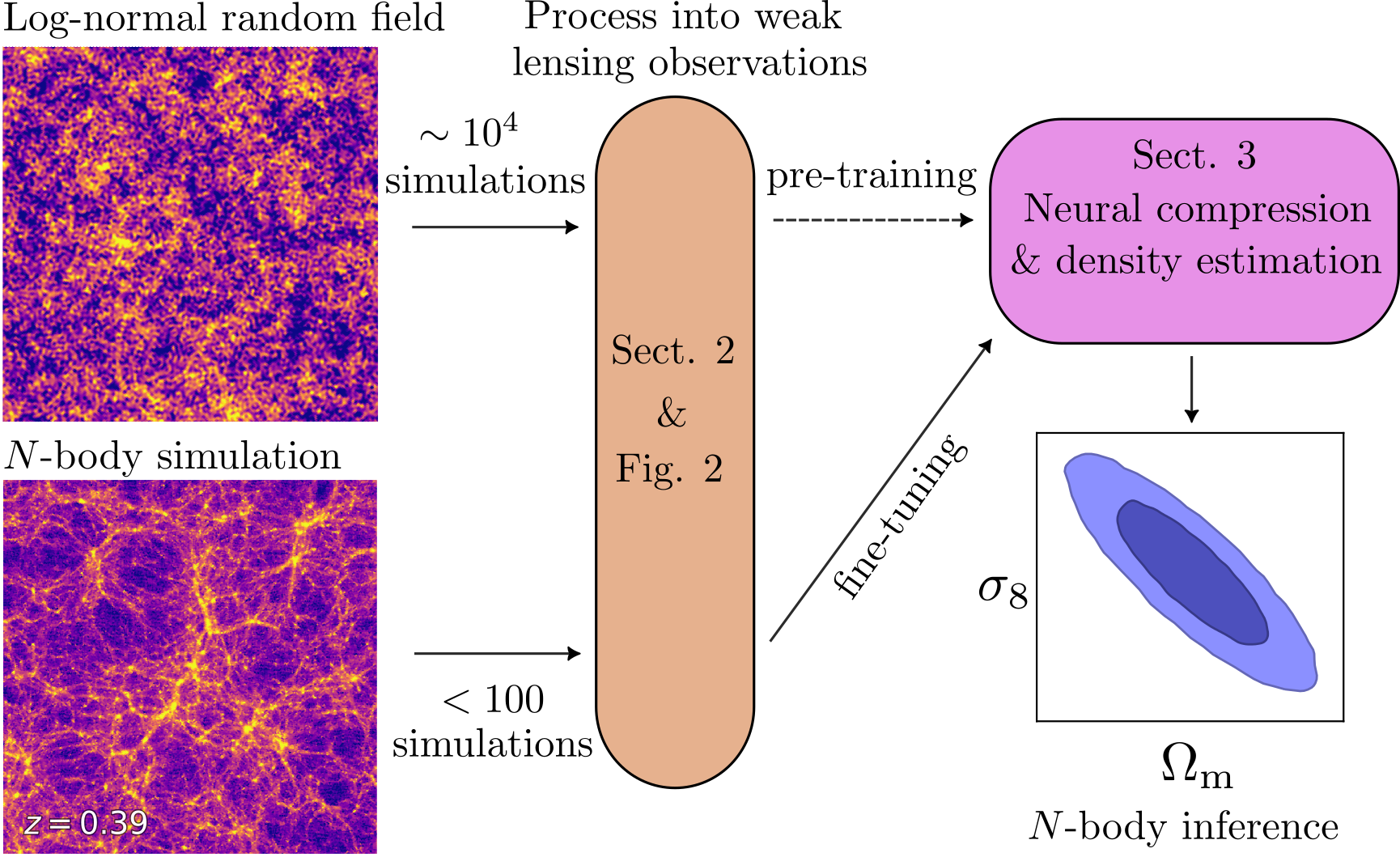}
    \caption{Our multifidelity SBI approach combines many cheap log-normal simulations (top left) with few expensive $N$-body simulations (bottom left), shown here as  $15^{\circ}\times15^{\circ}$ patches of the  matter overdensity fields $\log(1+\delta)$. These are processed into KiDS-Legacy-like weak lensing maps through a shared pipeline (described in Sect. \ref{sec:forward_modelling_kids}). We then leverage transfer learning, pre-training on the log-normal set and fine-tuning on the $N$-body set (described in Sect. \ref{sec:methodology}), to enable accurate and precise high-fidelity inference.}
    \label{fig:overview}
    
\end{figure}

Transfer learning is a natural framework for multifidelity SBI because ML models trained on a large dataset of low-fidelity simulations tend to learn generic and transferable features \citep{he2016deep, zhai2022scaling, krouglova2025multifidelity}. These learned representations can then be rapidly adapted to higher-fidelity domains using far fewer simulations. \citet{saoulis2025transfer} demonstrated this idea for cosmological SBI using the CAMELS Multifield Dataset \citep{villaescusa2022camels}, showing that neural compression and density estimation models pre-trained on lower-fidelity simulations could be adapted to higher-fidelity data with very few simulations. However, that work considered idealised simulation fields rather than realistic weak lensing observations.

Here, we demonstrate the feasibility and benefits of applying multifidelity SBI to cosmic shear. An overview of our approach is shown in \cref{fig:overview}. We produce KiDS-Legacy-like mock weak lensing observations at two fidelities. The lower fidelity uses log-normal random field simulations of the underlying matter power spectrum, while the higher fidelity set uses matter shells from the $N$-body Gower Street simulation suite \citep{jeffrey2025dark}. We design a suitable deep neural compression architecture to jointly compress the power spectrum and convergence maps using a hybrid learning scheme \citep{lucas2025hybrid}. We  pre-train this neural compression model, as well as neural density estimation models, on the lower fidelity log-normal simulations. We then explore how this pre-training enables accurate neural posterior estimation \citep[NPE,][]{papamakarios2016fast} on the high-fidelity $N$-body simulations even with very few simulations. We also present a scheme to perform multifidelity neural likelihood estimation \citep[NLE,][]{lueckmann2019likelihood, papamakarios2019sequential} for the first time, decoupling the high-fidelity simulation suite prior from the analysis and enabling likelihood-based inference with few simulations. We further show that a simple linear whitening and truncation of the learned summaries further improves the sample-efficiency and calibration of multifidelity NLE.

This manuscript is structured as follows. \Cref{sec:forward_modelling_kids} describes the simulation-based approach to forward model realistic, KiDS-Legacy-like observations that are amenable to map-level neural compression. \Cref{sec:methodology} describes the multifidelity SBI methodology, our neural compression and density modelling approach, and the model evaluation techniques we use. \Cref{sec:results} presents the performance gains achieved by multifidelity SBI relative to standard single-fidelity inference as a function of the number of available simulations. Finally, we present a discussion of our results in \cref{sec:discussion} and our conclusions in \cref{sec:conclusions}.

\section{Forward modelling KiDS-Legacy}
\label{sec:forward_modelling_kids}

We present a proof-of-concept for performing field-level inference directly with high-fidelity cosmological simulations in the context of current and upcoming surveys. We therefore produce mock observations that approach the realism of current cosmological analyses, including important systematic effects and nuisance parameters in the analysis. 

We closely follow the KiDS-SBI modelling framework, previously applied to KiDS-1000 \citep{lin2023simulation,von2025kids}. We describe the key modelling choices in brief here, and defer to this prior work for the complete descriptions of the practicalities and mathematical modelling. We include key systematic effects to approach the realism of the KiDS-Legacy analysis \citep{wright2025kids}, including photometric redshift uncertainties and a halo-mass-dependent non-linear alignment model of intrinsic galaxy shapes (NLA-M).  Compared to the KiDS-1000 dataset, KiDS-Legacy increases the survey area by 34\% and, through improved photometric calibration, extends the source redshift limit to $z\simeq 2$, yielding a 3.5-fold increase in survey volume.

For our high-fidelity simulation suite, we use the gravity-only full-sky $N$-body Gower Street simulations, with details presented in \citet{jeffrey2025dark}. These simulate $\nu\textrm{-}w\textrm{CDM}$ cosmologies using \texttt{PKDGRAV3} \citep{potter2017pkdgrav3} with a box size of $L=1250\:h^{-1}\textrm{Mpc}$ and $1080^3$ particles. Particle locations and velocities are initialised using second-order Lagrangian perturbation theory and evolved under Newtonian gravity. Time-dependent snapshots from these simulations are then processed into lightcones (using box replication where necessary) to produce concentric matter density shells spaced across redshift $z$. These snapshots are publicly available (\citealp{jeffrey2025dark}, see Data Availability). 

For the low-fidelity simulation suite, we use the semi-analytic non-linear matter power spectrum model \texttt{HMcode-2020} \citep{mead2021hmcode} implemented by \texttt{CAMB} \citep{lewis2011camb}. We then utilise the Generator for LArge Scale Structure (\texttt{GLASS}, \citealp{tessore2023glass}) package to generate log-normal random fields of the dark matter overdensity contrast with the correct two-point statistics. 

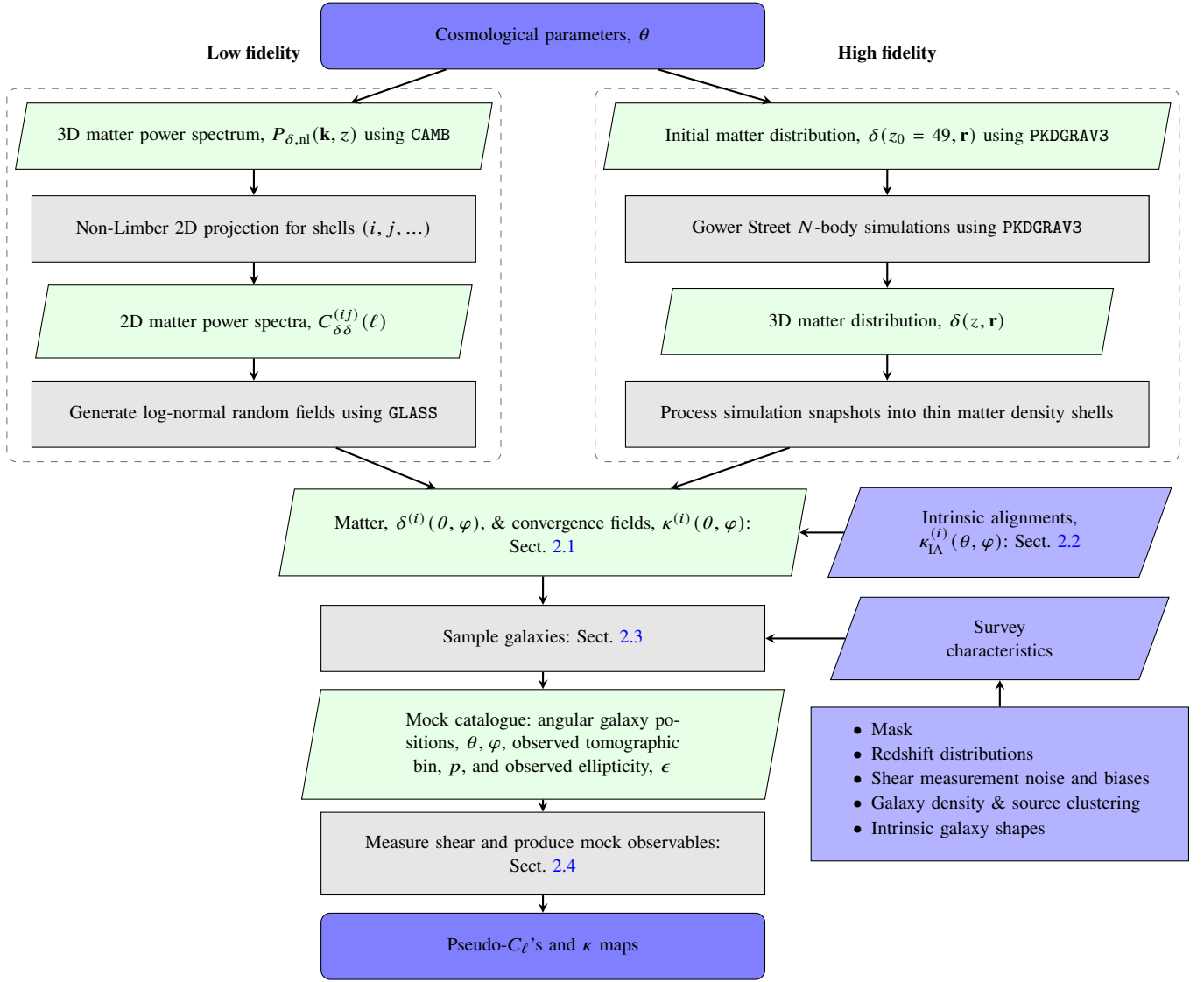
\begin{figure*}
    \centering
    \begin{tikzpicture}[node distance=1.4cm and 2.2cm]

        \node (start) [startstop, text width=6cm] {Cosmological parameters, $\theta$};

        \node (camb) [middle, below left=0.5cm and 0.5cm of start, text width=6cm]
        {3D matter power spectrum, $P_{\delta, \mathrm{nl}}(\mathbf{k}, z)$ using \texttt{CAMB}};

        \node (levin) [process, below of=camb]
        {Non-Limber 2D projection for shells $(i,j,...)$};

        \node (cl) [middle, below of=levin, text width=5.5cm]
        {2D matter power spectra, $C_{\delta \delta}^{(i j)}(\ell)$};

        \node (random) [process, below of=cl]
        {Generate log-normal random fields using \texttt{\texttt{GLASS}}};

    \node (initmatter) [middle,
        below=0.5cm of start,
        xshift=5.2cm,
        text width=7.2cm]
    {Initial matter distribution, $\delta(z_0=49, \mathbf{r})$ using \texttt{PKDGRAV3}};

    \node (nbody) [process, below of=initmatter, text width=7.2cm]
    {Gower Street $N$-body simulations using \texttt{PKDGRAV3}};

    \node (nbody) [process, below of=initmatter, text width=7.2cm]
    {Gower Street $N$-body simulations using \texttt{PKDGRAV3}};

        \node (nbodymatter) [middle, below of=nbody, text width=5.5cm]
        {3D matter distribution, $\delta(z, \mathbf{r})$ };

        \node (hshell) [process, below of=nbodymatter, text width=7.2cm]
        {Process simulation snapshots into thin matter density shells };

        \node (lowframe) [branchframe, fit=(camb)(random), label={[font=\bfseries]above:Low fidelity}] {};
        \node (highframe) [branchframe, fit=(initmatter)(hshell), label={[font=\bfseries]above:High fidelity}] {};

        \node (glass) [middle, text width=6.8cm] at (0,-7.5, 0)
        {Matter, $\delta^{(i)}(\theta, \varphi)$, \& convergence fields, $\kappa^{(i)}(\theta, \varphi)$: \\ Sect.~\ref{sec:weak_lensing}};

        \node (glass_in) [io, right of=glass, xshift=5.5cm, text width=3.5cm]
        {Intrinsic alignments, $\kappa_{\mathrm{IA}}^{(i)}(\theta, \varphi)$: Sect.~\ref{sec:intrinsic_alignments}};

        \node (sampling) [process, below of=glass, yshift=-0.2cm]
        {Sample galaxies: Sect.~\ref{sec:galaxy_sampling}};

        \node (salmo) [middle, below of=sampling, yshift=-0.2cm, text width=6cm]
        {Mock catalogue: angular galaxy positions, $\theta, \varphi$,  observed tomographic bin, $p$, and observed ellipticity, $\epsilon$};

        \node (salmo_in) [io, right of=sampling, xshift=5.5cm, text width=3.5cm]
        {Survey \\ characteristics};

        \node (salmo_in_in) [subio, below of=salmo_in, text width=5cm, yshift=-0.8cm, xshift=0cm]
        {\begin{itemize}\itemsep0.2em \vspace{-\baselineskip}
            \item Mask
            \vspace{-0.1\baselineskip}\item Redshift distributions
            \vspace{-0.1\baselineskip}\item Shear measurement noise and biases
            \vspace{-0.1\baselineskip}\item Galaxy density \& source clustering
            \vspace{-0.1\baselineskip}\item Intrinsic galaxy shapes
        \end{itemize}};

        \node (measure) [process, below of=salmo, yshift=-0.25cm]
        {Measure shear and produce mock observables: \\ Sect.~\ref{sec:mock_observations}};

        \node (pcl) [startstop, below of=measure, yshift=0cm, text width=6cm]
        {Pseudo-$C_\ell$'s and $\kappa$ maps};

        \draw [arrow] (start) -- (camb);
        \draw [arrow] (start) -- (initmatter);

        \draw [arrow] (camb) -- (levin);
        \draw [arrow] (levin) -- (cl);
        \draw [arrow] (cl) -- (random);

        \draw [arrow] (initmatter) -- (nbody);
        \draw [arrow] (nbody) -- (nbodymatter);
        \draw [arrow] (nbodymatter) -- (hshell);

        \draw [arrow] (random) -- (glass);
        \draw [arrow] (hshell) -- (glass);

        \draw [arrow] (glass) -- (sampling);
        \draw [arrow] (sampling) -- (salmo);
        \draw [arrow] (salmo) -- (measure);
        \draw [arrow] (measure) -- (pcl);
        \draw [arrow] (salmo_in_in) -- (salmo_in);

        \draw [arrow] (glass_in) -- (glass);
        \draw [arrow] (salmo_in) -- (sampling);

    \end{tikzpicture}
\caption{Flowchart describing the steps of our multifidelity simulation pipeline of cosmic shear observables from cosmological parameters. The dark blue rounded boxes represent the inputs and outputs of the simulation-based inference pipeline. The green slanted boxes represent relevant quantities which are calculated during the simulation. The grey rectangular boxes show steps in the calculations, while the blue slanted boxes show any (systematic) effects which are included. Indices $(i,j,\dots)$  label the matter shells used to discretise the line-of-sight integrals. }
\label{fig:sim_diagram}
\end{figure*}

The key steps of our multifidelity simulation approach are summarised in \cref{fig:sim_diagram}. We describe the details of our forward modelling framework below. 

\subsection{Weak lensing}
\label{sec:weak_lensing}

Weak lensing describes the bending of light from a distant source by the gravitational effect of the intervening matter density field. The gravitational potential, $\Phi$, is related to the matter overdensity $\delta\equiv\delta\rho/\overline{\rho}$ by:
\begin{equation}
    \nabla^2_r\Phi(t, \mathbf{r}) = \frac{3\Omega_\mathrm{m}H_0^2}{2a(t)}\delta(t, \mathbf{r}), 
\label{eq:gravitational_potential}
\end{equation}
for time $t$ and comoving spatial coordinate $\mathbf{r}$. $H_0$ is the present day Hubble constant, $\Omega_\mathrm{m}$ is the matter density parameter, and $a(t)=1/(1+z(t))$ represents the scale factor. We may then apply the Born approximation to the lens equation to estimate the lensing potential $\phi$ induced by this gravitational potential. For comoving distance $\chi$ and celestial sphere angles $(\theta, \varphi)$, we may integrate along the line-of-sight to each source galaxy, and average this over some arbitrary redshift distribution $n(z)$ of the source galaxies, giving:
\begin{equation}
    \phi(\theta,\varphi) = \frac{2}{c^2}\int_0^\infty \textrm{d}\chi \:n(z(\chi))\int^\chi_0 \textrm{d}\chi' \frac{(\chi -\chi')}{\chi\chi'}\Phi(\chi', \theta, \varphi).
\end{equation}
We have assumed curvature $K=0$. The induced convergence $\kappa$ is related to the second derivatives of this lensing potential \mbox{$\kappa=\nabla_r^2\phi/2$}; this allows for the convergence to be related to the overdensity  \citep{schneider2006weak, jeffrey2021dark}:
\begin{equation}
    \kappa(\theta,\varphi) = \frac{3\Omega_\mathrm{m}H_0^2}{2c^2}\int_0^\infty \textrm{d}\chi \:n(z(\chi))\int^\chi_0 \textrm{d}\chi' \frac{\chi'(\chi -\chi')}{\chi} \frac{\delta(\chi', \theta, \varphi)}{a(\chi')}\ .
    \label{eq:convergence_lensing_integrals}
\end{equation}
The shear field $\gamma$ can most straightforwardly be computed from the convergence field in the spherical harmonic basis. For spin-weighted $s$ spherical harmonic coefficients ${}_{s}Y_{\ell m}$
\begin{equation}
\kappa(\theta,\varphi) = \sum_{\ell m} \kappa_{\ell m} {}_{0}Y_{\ell m}(\theta,\varphi), 
\qquad
\gamma(\theta,\varphi) = \sum_{\ell m} \gamma_{\ell m} \, {}_{2}Y_{\ell m}(\theta,\varphi),
\end{equation}
where $\kappa_{\ell m}$ and $\gamma_{\ell m}$ are the spherical harmonic coefficients of the convergence and shear fields, respectively. Shear can be related to the convergence field by:
\begin{equation}
\gamma_{\ell m} = -\sqrt{\frac{(\ell + 2)(\ell - 1)}{\ell(\ell + 1)}} \, \kappa_{\ell m},
\label{eq:shear_convergence_harmonics}
\end{equation}
after which shear can be transformed back to real space through the discrete inverse spherical harmonic transform \citep{tessore2023glass, von2025kids}. 

This prescription relates shear to the underlying overdensity field, which can be generated either via theory or simulation. Concretely, we use \texttt{GLASS} to estimate the integrals in \cref{eq:convergence_lensing_integrals} via thin shell discretisation over many 2-D matter overdensity slices $\delta(\chi)$. Producing convergence and shear fields therefore requires some way of computing the 2-D matter overdensity at a given redshift.

For our multifidelity approach, we utilise two schemes for producing realisations of the overdensity $\delta(\chi)$ for a given cosmology. In the low-fidelity case, we use \texttt{HMcode-2020} to calculate the 3-D matter power spectrum and perform non-Limber integration with \texttt{CAMB} to produce 2-D spectra over 30 shells. These shells are evenly spaced over comoving distance between $z\in[0,2.0]$ in steps of 200 Mpc, producing around 20 to 24 shells depending on the cosmology. We use \texttt{GLASS} to produce log-normal overdensity fields for each shell at a resolution of $\mathtt{NSIDE}=1024$. This \textrm{HEALPix} resolution enables computation of the lensing power spectrum down to the desired smallest scale of $\ell_\textrm{max}=1500$. Within each shell we use linear integration windows to compute its contribution to the convergence \citep[i.e. performing the second integral in \cref{eq:convergence_lensing_integrals}, see][for a discussion of practicalities]{tessore2023glass}.

For the high-fidelity case, the $N$-body Gower Street simulations provide (around 75) thin matter shells on the celestial sphere between $z\in[0,2.0]$. We first convert them to overdensity shells and degrade them to $\mathtt{NSIDE}=1024$. These are then passed to \texttt{GLASS} with tophat window functions to perform the per-shell overdensity integrations over redshift $z$. This procedure for producing cosmology and redshift-dependent overdensity fields is the only difference between the two fidelities, as shown in \cref{fig:sim_diagram}. 

\subsection{Intrinsic alignments}
\label{sec:intrinsic_alignments}

Galaxy formation is influenced by the surrounding gravitational environment, which can lead to non-random, intrinsically aligned galaxies. Left untreated, such effects can bias cosmological inference even for two-point statistics at Stage-III survey precision \citep{troxel2015intrinsic, krause2016impact}. Explicit treatment of intrinsic alignments is therefore essential when using higher-order statistics. 

Theoretical, numerical, and observational evidence indicates that these alignments will depend on both the galaxy type, with only early-type galaxies experiencing significant non-zero intrinsic alignments, and on the host halo mass \citep{catelan2001intrinsic, hirata2004intrinsic, hirata2007intrinsic, piras2018mass, fortuna2021kids}. We therefore select  NLA-M as introduced in \citet{wright2025kids} as a representative intrinsic alignments model, which modulates the intrinsic alignment signal depending on the red galaxy fraction $f_r$ and the average halo mass $\overline{M}_h$ at a given redshift. We may directly compute the impact of intrinsic alignments as an additive correction to the convergence $\kappa(z)$ through the semi-analytic model:
\begin{equation}
    \kappa_\textrm{IA} (z, \theta, \varphi) = -A_\textrm{IA} f_r^{(i)}\left(\frac{\overline{M}_h^{(i)}}{M_\textrm{h,piv}}\right)^{\beta_\textrm{IA}}\frac{C_1\rho_\mathrm{cr}\Omega_\mathrm{m}}{D(z)}\delta(z, \theta, \varphi).
\end{equation}
Here, $C_1 \equiv 5\times10^{-14} (h^2 \mathrm{M}_\odot /\textrm{Mpc}^{-3})^{-2}$ is an effective normalisation that sets the amplitude of the intrinsic alignment signal \citep{bridle2007dark}, $\rho_\mathrm{cr}$ is the critical density of the Universe today, $D(z)$ is the linear growth factor normalised to unity at $z=0$, and $M_\textrm{h,piv} = 10^{13.5}h^{-1}\mathrm{M}_\odot$ is the pivot halo mass scale. The red galaxy fraction $f_r^{(i)}$ in each tomographic bin, estimated from KiDS photometric data, was found to be $\approx 0.15$ to $0.2$ in the first five bins but negligible in the sixth. The average halo mass $\overline{M}_h^{(i)}$ is derived from KiDS DR5 $r$-band luminosities through an empirical luminosity- and redshift-dependent mass relation, ranging from $\log_{10}\overline{M}_h \approx 11.7$ to $13.2$ across the six bins; we defer to \citet{wright2025kids} for further details. $A_\textrm{IA}$ and $\beta_\textrm{IA}$ are free coefficients that can be determined empirically. 

Following \citet{wright2025kids}, we use the joint posterior found on $\{A_\textrm{IA}, \beta_\textrm{IA}\}$ in \citet{fortuna2025kids} as a starting point for our analysis. For dataset generation, we use uniform priors on the intrinsic alignment parameters $A_\mathrm{IA} \sim\mathcal{U}[4.48,7.0]$ and $\beta_\mathrm{IA} \sim\mathcal{U}[0.28,0.6]$, ensuring this covers the results of \citet{fortuna2025kids} at $5\sigma$.

It is worth noting that these priors are relatively informative because they are calibrated from a dedicated KiDS galaxy–galaxy lensing measurement that directly constrains the IA-halo-mass relation \citep{fortuna2025kids}. This contrasts with analyses based on model-agnostic intrinsic alignment prescriptions such as non-linear alignments (NLA), which typically adopt substantially broader priors. We therefore do not expect KiDS-Legacy weak lensing observations to produce significant constraints on these parameters beyond the prior. 

\subsection{Galaxy sampling and survey systematics}
\label{sec:galaxy_sampling}

\subsubsection{Source clustering}

Galaxies preferentially trace the underlying matter density distribution, leading to source clustering. While two-point weak lensing analysis is not strongly impacted by this effect \citep{krause2021dark, wright2025kids}, higher-order statistics (and therefore field-level observations) are more sensitive to source clustering \citep{gatti2024dark, jeffrey2025dark}. We model source clustering using a linear galaxy bias prescription.  The correction takes the form of:
\begin{equation}
n(z, \theta, \varphi)
=
\bar n(z)\left[1 + b_g \delta(z, \theta, \varphi) \right],
\end{equation}
where $\bar n(z)$ is the sky-averaged redshift distribution and $b_g$ is the linear galaxy bias parameter. Throughout this work we fix $b_g=1$, a reasonable approximation for the deep, faint source samples used here, whose effective large-scale linear bias is of order unity \citep{jullo2012cosmos, gatti2024dark}. For application to real data in future work, this can be allowed to vary. Galaxies are then sampled from each shell according to the local effective galaxy density.

\subsubsection{Photometric redshift uncertainties}

We adopt the KiDS-Legacy gold sample photometric redshift distributions for each tomographic bin $i$, $\bar n^{\mathrm{gold}}_i(z)$  from \citet{wright2025kids}. 

In order to account for potential biases in these inferred distributions, an additional random shift with a given mean and covariance for each bin is performed. This random shift uses the uncertainties derived from a calibration step that leverages the semi-analytic galaxy imaging simulations SKiLLS \citep{li2023kids} in \citet{wright2025kidsredshift}.  For a given simulation, we therefore sample a different randomly shifted photo-$z$ distribution:
\begin{equation}
    \bar n_i(z)
    =
    \bar n^{\mathrm{gold}}_i(z - \delta z_i),
    \quad
    \boldsymbol{\delta z}
    \sim
    \mathcal{N}(\boldsymbol{\mu}_{\delta z}, \mathbf{C}_{\delta z}),
\end{equation}
where we adopt the KiDS-Legacy estimates of the mean photo-$z$ shifts $\boldsymbol{\mu}_{\delta z}$ and their covariance $\mathbf{C}_{\delta z}$ \citep{wright2025kids}.

We use the KiDS-Legacy effective galaxy number densities for each tomographic bin,
$\bar{n}_{\mathrm{eff}, i}^{\mathrm{gold}} = (1.77,\: 1.65,\: 1.50,\: 1.46,\: 1.35,\: 1.07)\,\mathrm{arcmin}^{-2}$, which set the total galaxy density associated with each shifted redshift distribution $\bar n_i(z)$. For each shell in the line-of-sight integration of \cref{eq:convergence_lensing_integrals}, we estimate the expected number of galaxies to be assigned to each tomographic bin. We then Poisson sample galaxies within each shell according to the effective galaxy density per-tomographic bin \citep{tessore2023glass}.

\subsubsection{Galaxy shape noise and biases}
\label{sec:shape_noise_annd_bias}

We model the noise of the underlying galaxies, comprising both intrinsic variability of the galaxy ellipticity and measurement noise, by randomly sampling galaxy shapes with \texttt{GLASS}. Galaxy ellipticity is a complex quantity, $\epsilon = \epsilon_1 + i\epsilon_2$, with the two components indexed by $k \in \{1, 2\}$. We use the KiDS-Legacy estimates for this intrinsic galaxy shape dispersion $\sigma_\epsilon$ per-component which takes slightly different values in each tomographic bin:
\begin{equation}
    \sigma_\epsilon = [0.277, 0.272, 0.290, 0.262, 0.280, 0.300].
\end{equation}

After producing lensed ellipticities for each galaxy $\epsilon^\mathrm{lensed}_k$, we also apply the multiplicative and additive biases derived from KiDS-legacy. These are footprint dependent, i.e. different for the north and south patches of the KiDS footprint. Concretely, for each mock observation we randomly sample a multiplicative and additive shear bias using the KiDS-Legacy estimates in \citet{wright2025kids}. The observed galaxy ellipticity $\epsilon^\mathrm{obs}_k$  is then modelled as:
\begin{equation}
    \epsilon^\mathrm{obs}_k = (1 + m) \epsilon^\mathrm{lensed}_k + c_k,
\end{equation}
for per-footprint multiplicative shears $m$ and additive shears $c_k$.

 \begin{figure*}
    \includegraphics[width=\textwidth]{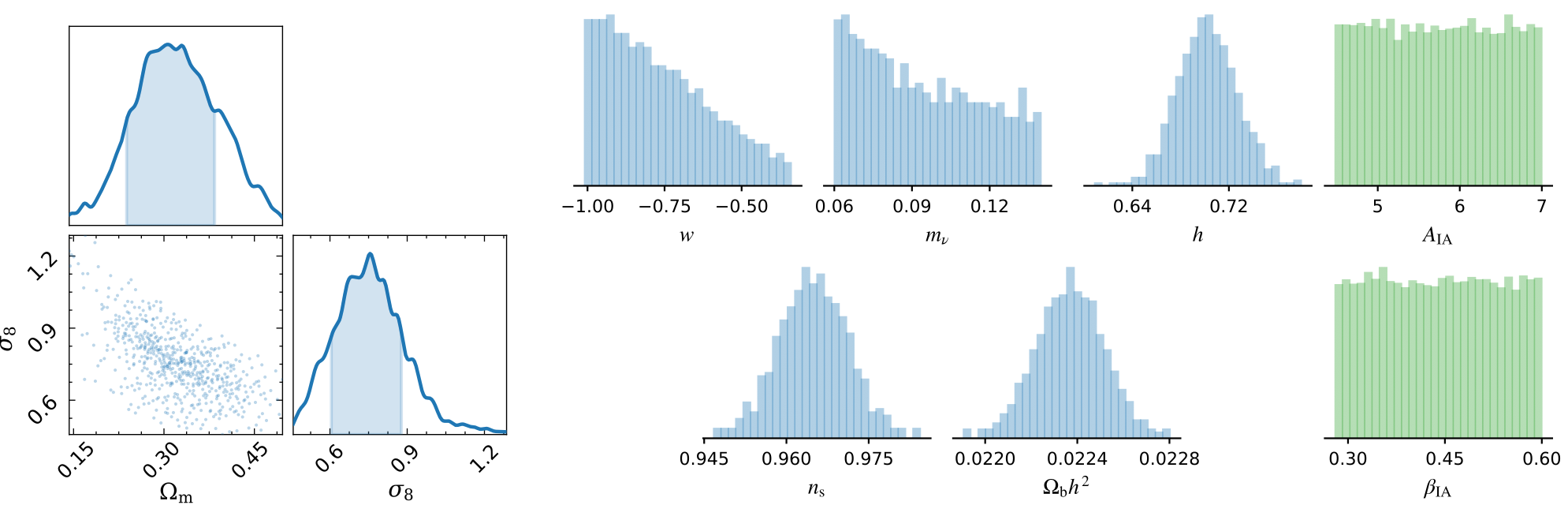}
    \caption{Cosmological parameter distributions of the Gower Street simulations (blue densities and samples), together with the intrinsic alignment parameter priors adopted in this work (green densities). The $\sigma_8$ and $\Omega_\mathrm{m}$ parameters were initially sampled with a joint prior distribution and then augmented with active learning. All other parameters were drawn from 1-D independent distributions; see \citet{jeffrey2025dark} for precise details.}
    \label{fig:gower}
    
\end{figure*}

\subsection{Mock observations}
\label{sec:mock_observations}

Mock observations are produced by simulating KiDS-Legacy-like galaxy catalogues with all the above systematics forward modelled. We use $\mathtt{NSIDE}=1024$ and perform all galaxy sampling on the \textrm{HEALPix} sphere, with the fractional KiDS-Legacy mask applied during sampling. The resulting catalogues are then processed into summaries that will serve as our mock observations.

We first process the catalogue into pixelised shear fields per tomographic bin $i$, correcting for both multiplicative and additive shear biases. For each \textrm{HEALPix} pixel $p$, we compute the shear field as
\begin{equation}
    \epsilon^{(p)}_{k,i} = \frac{1}{1+\overline{m_i}} \, \frac{\sum\limits_{g \in p} w_g \left( \epsilon^\mathrm{obs}_{k,g} - \langle \epsilon^\mathrm{obs}_k \rangle_i \right)}{\overline{W}_i} \,, \qquad \overline{W}_i \equiv \frac{1}{N_{\mathrm{pix}}} \sum_{g \in i} w_g \,,
\end{equation}
where the sum in the numerator runs over all galaxies $g$ falling within pixel $p$, and $\overline{W}_i$ is the mean galaxy count per pixel in tomographic bin $i$.\footnote{Normalising by the (mean) pixel occupancy rather than by the per-pixel weight sum avoids dividing the map by a stochastic quantity and yields simpler, better-behaved estimator statistics, following \citet{hall2025pixelization}.} The quantity $w_g$ is the lensing weight assigned to each galaxy (e.g.\ from shape measurement uncertainties) and $\langle \epsilon_k \rangle_i$ is the mean ellipticity in tomographic bin $i$, which corrects for additive shear biases. For the mock catalogues considered here, we set $w_g = 1$ for all galaxies, such that the estimator reduces to the per-pixel summed ellipticity divided by the mean galaxy count per pixel, $\overline{W}_i \rightarrow \bar{N}_i = N_{\mathrm{gal},i} / N_{\mathrm{pix}}$. The factor $\overline{m_i}$ denotes the mean multiplicative shear bias in tomographic bin $i$, and is applied as a global calibration factor. 

We produce pseudo-$C_\ell$ measurements by computing the auto- and cross-power spectra across all combinations of tomographic bins using \texttt{anafast}. We perform scale-cuts on the pseudo-$C_\ell$'s between $\ell = [76, 1500]$. We discard the B-modes and do not perform any unmixing, since the E/B-mode leakage induced by the KiDS mask has been shown to be negligible over this multipole range \citep{loureiro2022kids}. These pseudo-$C_\ell$'s are then processed into bandpowers with 8 bins for each power spectrum measurement using the \citet{brown2005cosmic} binning scheme \citep[see][for a more detailed treatment, which we follow]{von2025kids}. This produces a bandpower vector of dimension (21, 8) for the 6 tomographic bin combinations ($N\times(N+1)/2 = 21$ auto- and cross-spectra $\times$ 8 bandpower bins).  

For field-level cosmological information, we process the per-tomographic bin measured shear fields into convergence maps using the spherical harmonic relation in \cref{eq:shear_convergence_harmonics} (i.e. inverse Kaiser-Squires, see e.g. \citealt{jeffrey2021dark, wallis2021mapping}). We apply a Gaussian filter with full-width half-maximum of $8.0$ arcmin to remove field-level information at smaller scales that may contain mismodelling issues, and downsample the resulting convergence fields onto a HEALPix sphere of $\mathtt{NSIDE}=512$ (which resolves our smallest retained scales while improving computational efficiency). Finally, we pixelise the convergence maps by recentering each KiDS footprint onto the celestial equator and nearest-neighbour resampling the HEALPix maps onto a Cartesian grid. This produces field-level observations of approximate size (1000, 100) pixels for each of the 6 tomographic bins for both KiDS-North and KiDS-South. 

\subsection{Dataset generation}
\label{sec:dataset_generation}

 \begin{figure}
    \includegraphics[width=\columnwidth]{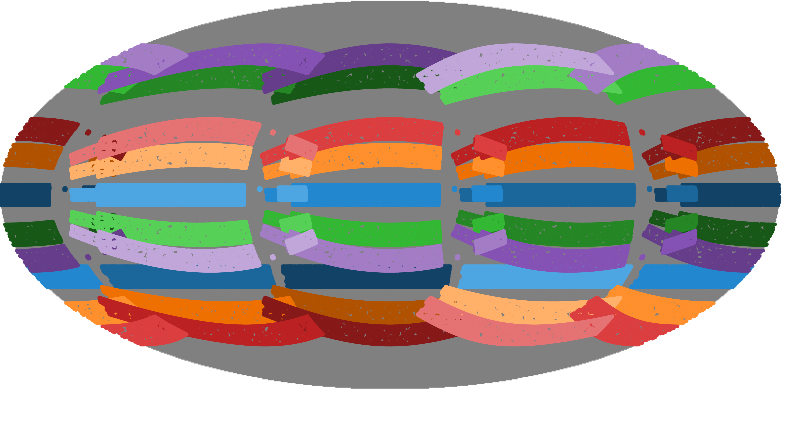}
    \caption{Twenty mostly disjoint KiDS-Legacy-like footprints on the sky, each coloured differently (with pairs of same colour patches corresponding to the North and South footprints). We use these as cutouts of the $N$-body Gower Street simulation lightcones to augment the number of mock observations with pseudo-independent underlying matter distribution realisations. }
    \label{fig:kids_footprints}
    
\end{figure}

We produce large suites of mock observations at two different fidelities: one using log-normal simulations with \texttt{GLASS}, and the other using the Gower Street $N$-body simulation suite. 

We describe the high-fidelity $N$-body suite first. The Gower Street simulations varied seven cosmological parameters $\{\Omega_\mathrm{m}, \sigma_8, w, m_\nu,  h, n_\mathrm{s}, \Omega_\mathrm{b}\}$. The simulation prior, shown in \cref{fig:gower}, is non-uniform for all parameters. In addition, a subset of the simulations were run with $\Omega_\mathrm{m}$ and $\sigma_8$ selected using active learning, so the effective Gower Street prior has no clean analytic form.

For the $600$ large-box cosmological simulations in the main Gower Street suite (excluding 191 ``science verification'' runs), we produce 20 KiDS-Legacy-like footprints of the underlying matter density fields. These footprints are shown in \cref{fig:kids_footprints}. This is achieved by rotating the underlying matter density fields in spherical harmonic space to produce KiDS footprints that are largely disjoint (we do not rotate the masks as the hard boundaries cannot be preserved at finite $\ell$ in the spherical harmonic space). For each rotated footprint of the underlying matter shells, we also produce four different realisations of the shape noise and systematic effects with varying nuisance parameters.  

This produces a heavily augmented dataset with 80 pseudo-independent KiDS-Legacy-like mocks per cosmological simulation. Note that the underlying matter density lightcones are also pseudo-independent since the large $N$-body box is replicated multiple times to produce the matter density shells \citep{jeffrey2025dark}. Non-independence is not a major concern here as each simulation is a valid draw from the forward model; correlated samples reduce the effective dataset size, but any degradation in accuracy (through for instance memorisation of features in a given simulation box) can always be detected and diagnosed on a held-out validation set. In total, our dataset generation procedure produces roughly 47,000 high-fidelity mock observations at 590 cosmologies (where 10 cosmologies failed due to issues with \texttt{CAMB}). 

There is considerable flexibility in producing the lower fidelity dataset. While in principle our multifidelity approach can handle differing priors at the different fidelities, for simplicity we roughly match the priors over both cosmological and nuisance parameters to the high-fidelity Gower Street suite. To produce our \texttt{GLASS} log-normal mocks we empirically model each 1-D marginal and sample from the prior of each cosmological parameter independently. 

We compute matter power spectra using \texttt{CAMB} for approximately 6500 cosmologies. For each cosmology, we generate five independent realisations of the nuisance and systematic parameters. For each resulting cosmology--nuisance parameter combination, we then generate four independent realisations of the log-normal fields and perform galaxy sampling. This yields up to 20 mock observations per cosmology, reducing computational costs by minimising the number of times \texttt{CAMB} must be run. We repeated this procedure until reaching a low-fidelity dataset with around 100,000 mock observations. Following prior multifidelity work \citep{jia2024cosmological,krouglova2025multifidelity, saoulis2025transfer}, this yields over an order of magnitude more low-fidelity simulations than the small subsets of the Gower Street suite used as the high-fidelity dataset.

\begin{figure}
    \includegraphics[width=\columnwidth]{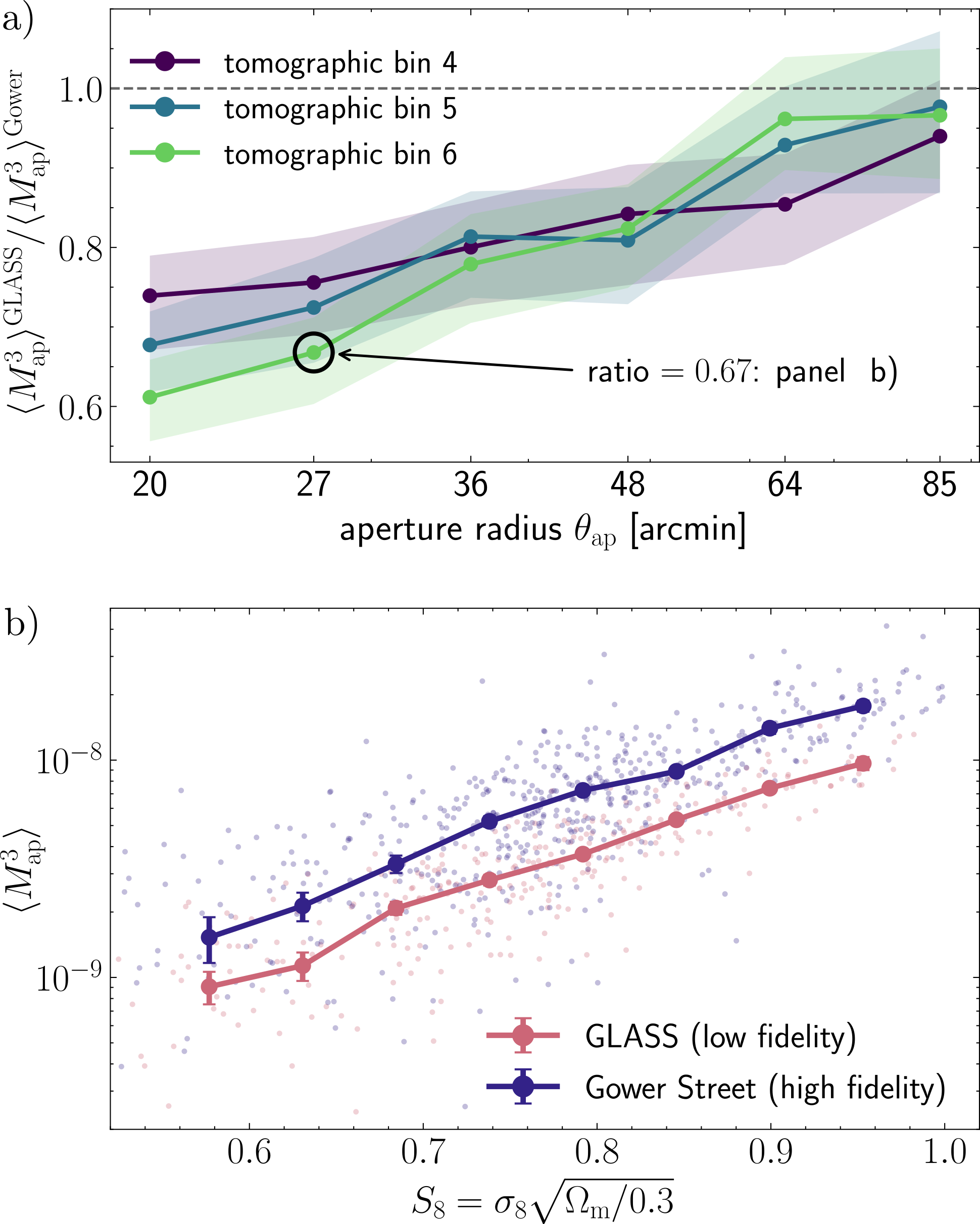}
    \caption{Differences in the third moment of the aperture mass, $\langle M_\mathrm{ap}^3 \rangle$ between the two fidelities. Panel a) shows the  $\langle M_\mathrm{ap}^3 \rangle$ as a function of smoothing scale and redshift ratio, averaged across the datasets and controlling for $S_8$. Panel b) shows the distribution of the aperture mass statistics for a fixed aperture radius across the two fidelities as a function of $S_8$.}
    \label{fig:fidelity_comparison}
    
\end{figure}

We present a comparison of the $\langle M_\mathrm{ap}^3 \rangle$ statistics of the two fidelities in \Cref{fig:fidelity_comparison}. While the low signal-to-noise ratio of the KiDS convergence maps makes the differences between the two fidelities imperceptible by eye, there are significant differences in the statistics of the final convergence maps---for instance, in the 1-point function of the maps, and even in the 2-point statistics (due to source clustering; not shown). We provide a brief discussion on the sensitivity of our multifidelity approach to the difference between fidelities in \Cref{sec:discussion}.

\section{Method}
\label{sec:methodology}

We aim to perform field-level inference with a highly restricted set of high-fidelity simulations. We therefore design a neural compression architecture that can process field-level statistics from \cref{sec:dataset_generation} to produce informative summaries, and can be efficiently adapted to a few high-fidelity examples. We utilise transfer learning, whereby a deep compression network and normalising flow are first pre-trained on the large low-fidelity dataset. These models are then fine-tuned (i.e. trained for a short period with smaller weight updates) on the much smaller set of high-fidelity examples. Prior work has demonstrated transfer learning can be highly effective in an SBI setting for training both effective neural compression architectures and accurate density estimation models \citep{saoulis2025transfer, krouglova2025multifidelity, thiele2025simulation, hikida2025multilevel}. 

\subsection{Training techniques}

This work utilises multifidelity neural posterior estimation (NPE, \citealt{papamakarios2016fast}) and introduces an approach to perform multifidelity neural likelihood estimation (NLE). For all experiments in the main manuscript, we perform inference over 9 parameters: the 7 cosmological parameters varied in the Gower Street simulations, and the two intrinsic alignment parameters (see \cref{fig:gower}). The remaining nuisance parameters, such as those parametrising photometric uncertainties and shear biases, are prior-dominated and pre-marginalised in our framework. 

\subsubsection{Neural posterior estimation and optimal compression}

If one has control over the high-fidelity prior, NPE is appealing for its simplicity. More importantly, some form of NPE is convenient for SBI since it enables a simple procedure for training informative neural compression algorithms. 

Specifically, the objective of a compression algorithm is generally to maximise some measure of shared information between the compressed statistics and the cosmological parameters. One popular approach is Variational Mutual Information Maximisation (VMIM) which provides an informationally optimal guarantee on the learnt summaries \citep{jeffrey2021dark, lanzieri2024optimal}. Mutual information (MI) $I(t, \theta)$ between summaries $t$ and inference parameters $\theta$ can be expressed in terms of the Kullback-Leibler (KL) divergence $D_{KL}$ between the joint and the product of marginals:
\begin{equation}
    \begin{aligned}
    I(t, \theta) 
    &= D_{\mathrm{KL}}\big(p(t, \theta)\,\|\,p(t)p(\theta)\big)  \\
    &= \int d\theta \, dt \; p(t, \theta) \log \frac{p(t, \theta)}{p(t)p(\theta)} \\
    &= \int d\theta \, dt \; p(t, \theta) \log \frac{p(\theta \mid t)}{p(\theta)} \\
    &= \int d\theta \, dt \; p(t, \theta) \log p(\theta \mid t)
     - \int d\theta \, dt \; p(t, \theta) \log p(\theta) \\
    &= \int d\theta \, dt \; p(t, \theta) \log p(\theta \mid t)
     - \int d\theta \; p(\theta) \log p(\theta) \\
    &= \mathbb{E}_{p(t,\theta)}[\log p(\theta \mid t)]
     - \mathbb{E}_{p(\theta)}[\log p(\theta)] \\
    &= \mathbb{E}_{p(t,\theta)}[\log p(\theta \mid t)] + H(\theta).
    \end{aligned}
\end{equation}
$H(\theta)$ is the entropy, which can be ignored during an optimisation algorithm as it is a constant for a fixed prior. This leaves just the left hand term to be optimised during training. Defining two neural networks: compressor $F_\phi\!:\!x\rightarrow t$ and sufficiently flexible density estimation head $q_\varphi$, we can define a training objective to be maximised variationally:
\begin{equation}
\max_{\phi, \varphi} \; \mathbb{E}_{p(x,\theta)} \big[ \log q_{\varphi}(\theta \mid F_{\phi}(x)) \big].
\label{eq:MI_lower_bound}
\end{equation}
This corresponds to the standard NPE objective; i.e. training a neural network to learn a model of the posterior. This has the convenient corollary of training a compressor $F_\phi$ to produce summaries $t$ that extract maximal cosmological information from the observables $x$. In addition, one can read off the loss of the trained model to estimate the MI of the compression algorithm \citep{sui2026evaluate}. We denote this quantity $\Delta$MI; the $\Delta$ indicates that it equals the mutual information up to a fixed additive constant.  

We utilise a hybrid learning approach to learn beyond-two-point information from the convergence fields \citep{lucas2025hybrid}. Hybrid learning trains compression models in successive stages, concatenating previously learned summaries at each stage. This encourages the compression networks to learn new, complementary information from the observations relative to earlier stages. 

\begin{figure*}
    \includegraphics[width=\textwidth]{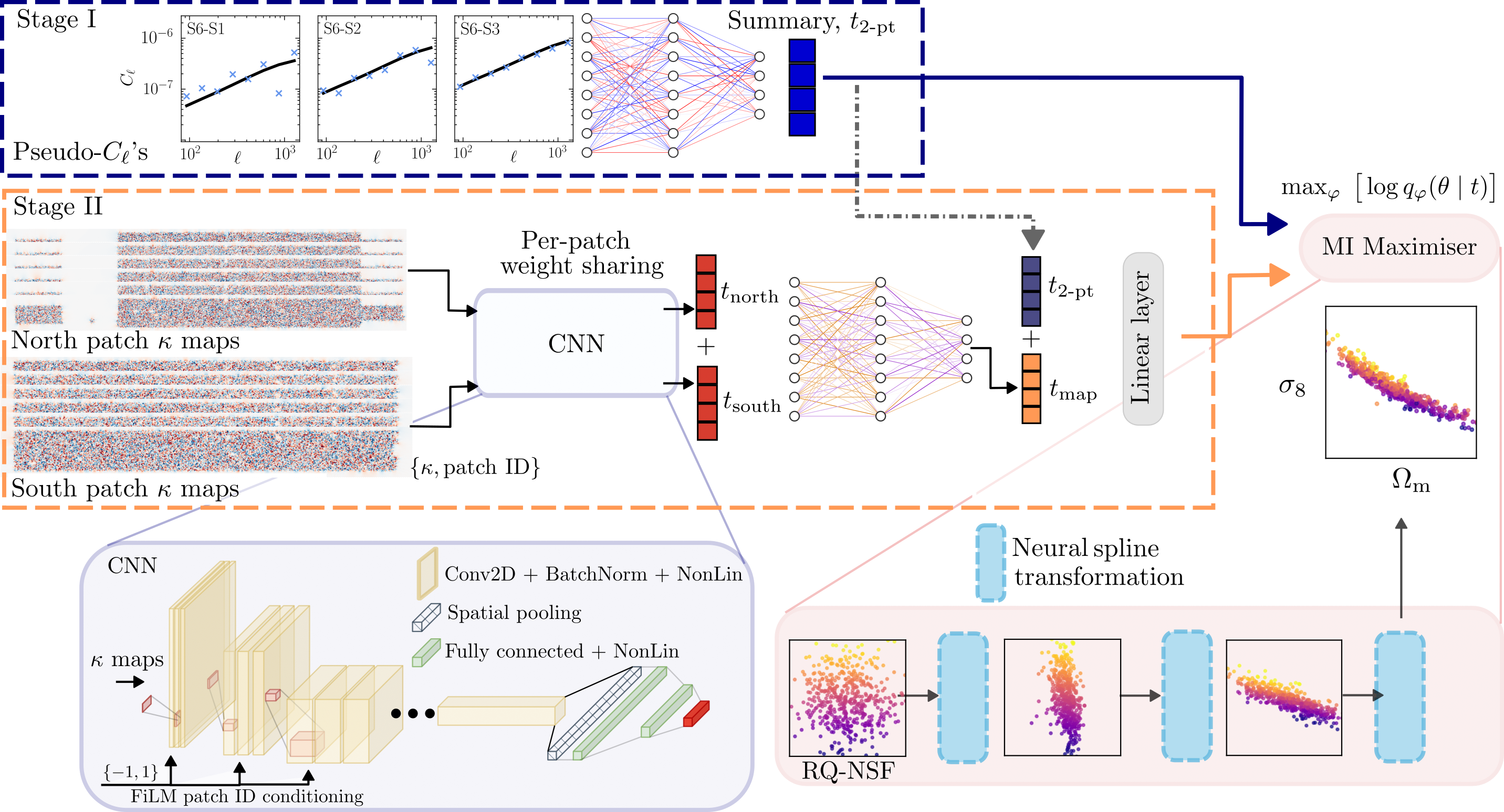}
    \caption{Our hybrid learning compression scheme to learn higher-order statistics from the convergence maps $\kappa$ using a CNN. In Stage I, a feedforward network is trained to compress the pseudo-$C_\ell$ measurements into summaries $t_\textrm{2-pt}$, after which the network is frozen. In Stage II, a CNN processes both the North and South KiDS patches into summaries $t_{\textrm{north},\textrm{south}}$. These are combined and further compressed into beyond 2-pt map-level statistics $t_\textrm{map}$, which are concatenated with the frozen 2-point embeddings to produce the final summary. We use a rational-quadratic neural spline flow for mutual information (MI) maximisation, which learns a model of the posterior $q_\varphi(\theta \mid t)$. The same trained compressor is then used to learn a likelihood model $q_\psi(t \mid \theta)$ for NLE (not shown). }
    \label{fig:hybrid_architecture}
    
\end{figure*}

We first train a compression algorithm on the bandpower observations with the VMIM objective to produce optimal summary statistics $t_\textrm{2-pt}$. During the map-level training stage, we freeze the bandpower compression network and concatenate $t_\textrm{2-pt}$ with the outputs of the map compressor $t_\textrm{map}$. This allows the field-level compressor to focus on extracting beyond-two-point information from the convergence maps. A schematic of our architecture and training approach is presented in \cref{fig:hybrid_architecture}.

\subsubsection{Neural likelihood estimation}

A major disadvantage of NPE is that it freezes in the prior (both over cosmological parameters and nuisance parameters) used to produce the training dataset. NLE, on the other hand, learns a surrogate likelihood of the forward model (generally over a compressed representation of the observables), and can therefore be combined with posterior sampling strategies like Markov chain Monte Carlo (MCMC) to enable inference with arbitrary priors. We first train a compressor $F_{\phi}$ with the VMIM objective of \cref{eq:MI_lower_bound} and freeze its weights to $\phi^*$. A likelihood model $q_\psi(t \mid \theta)$ is then trained on the frozen summaries $t = F_{\phi^*}(x)$ by maximising
\begin{equation}
    \max_{\psi} \; \mathbb{E}_{p(x,\theta)} 
    \big[ \log q_{\psi}\big(F_{\phi^*}(x) \mid \theta\big) \big].
\label{eq:NLE_objective}
\end{equation}
This is the NPE objective of \cref{eq:MI_lower_bound} with the arguments of the density estimator exchanged to model the likelihood.

In order to incorporate arbitrary priors $p(\theta)$, one must ensure sufficient sensitivity to parameters $\theta$ in the compressed representation $t$. This is best achieved by including in the VMIM training any parameters $\theta$ (both cosmological and nuisance) whose prior we wish to modify. Otherwise, low or zero sensitivity to these parameters in the compressed representation will lead to an effective prior that approaches the training prior, marginalising over unmodelled inference parameters. This is discussed in e.g. \citet{jeffrey2025dark}, where most nuisance parameter priors were not modified from the Gower Street prior. 

\subsubsection{Multifidelity approaches}
\label{sec:multidelity_approaches}

We explore several approaches for utilising the lower-fidelity log-normal simulations to enable high-fidelity inference with a limited number of simulations. These all involve pre-training both the compression and density estimation architectures, before fine-tuning on a limited set of high-fidelity examples. Common to all these is pre-training the compression and inference models $F_\phi$ and $q_\varphi$ using the hybrid learning scheme on the full dataset of \texttt{GLASS} log-normal simulations. We denote the pre-trained model weights as $F_{\phi^*}$ and $q_{\varphi^*}$. We explore the following schemes:
\begin{enumerate}
    \item \textit{NPE}: Finetune NDE only; freeze $F_{\phi^*}$ and finetune the pre-trained NDE head weights $q_{\varphi^*}$ on the high-fidelity data.
    \item \textit{NPE}: Finetune all; finetune all pre-trained weights from both the compressor $F_{\phi^*}$ and NDE head $q_{\varphi^*}$.
    \item \textit{NLE}: Finetune NLE; freeze $F_{\phi^*}$ and pre-train an NLE model $q_{\psi^*}(t\mid \theta)$ on the low-fidelity dataset. Then, compress the high-fidelity dataset with the same frozen compressor $F_{\phi^*}$ and fine-tune the pre-trained NLE $q_{\psi^*}(t\mid \theta)$ on the high-fidelity simulations.
\end{enumerate}
We note that items (i) and (ii) have been explored in prior work \citep[e.g.,][]{saoulis2025transfer, thiele2025simulation}. The multifidelity NLE approach in (iii) is demonstrated for the first time in this work. Notably, none of our approaches require the pairing of cosmological parameters or initial conditions between fidelities \citep[in contrast to other prior works, e.g.][]{chartier2021carpool,lee2024zooming, thiele2025simulation, hikida2025multilevel}. 

\subsubsection{Whitening and truncating the summary space for NLE}
\label{sec:summary_whitening}

Prior work has noted that mutual information maximisation does not necessarily produce representations that are well-behaved: the summaries may have a highly complex joint distribution, with strongly degenerate summaries and significant non-Gaussianity \citep{poole2019variational, Tschannen2020On}. This is of particular concern for NLE, which must learn to model the full distribution over $t$ (as opposed to NPE, which is just conditioned on it). A highly complex likelihood structure $p(t \mid \theta)$ is difficult to model and may increase the number of simulations required to ensure calibration \citep{ganem2022diagnosing,makinen2026degeneracy}. There are many principled approaches to mitigate this issue at training time. However, we found that applying several of these techniques such as regularising the latent space towards an approximately Gaussian distribution (as in a variational autoencoder bottleneck), or the using VICReg objective \citep{bardes2022vicreg}, did not improve the downstream high-fidelity NLE performance. 

Instead, we find that rotating the embedding space into an orthonormal frame using principal component analysis (PCA), and keeping a reduced number of components, is sufficient to substantially improve performance. Let $\{t_i\}_{i=1}^{N_{\rm LF}}$ be the summaries obtained by applying the frozen compressor $F_{\phi^*}$ to the low-fidelity \texttt{GLASS} training set, with sample mean and per-dimension sample standard deviations $\sigma_1, \dots, \sigma_{d}$. We first standardise each summary dimension, writing $\mathbf{D} = \mathrm{diag}(\sigma_1, \dots, \sigma_{d})$, so that no single dimension dominates the decomposition purely by virtue of its scale. We then diagonalise the covariance of the standardised summaries, $\mathbf{C} = \mathrm{Cov}\left[\mathbf{D}^{-1}(t - \bar{t})\right]$ (equivalently, the correlation matrix of $t$), as $\mathbf{C} = \mathbf{U} \mathbf{\Lambda} \mathbf{U}^{\mathrm{T}}$, where $\mathbf{\Lambda} =\mathrm{diag}(\lambda_1, \dots, \lambda_{d})$ with $\lambda_1 \geq \dots \geq\lambda_{d}$ eigenvalues and the columns of $\mathbf{U}$ are the corresponding eigenvectors. Retaining the leading $k \leq d$ components, we write $\mathbf{U}_k \in\mathbb{R}^{d \times k}$ for the first $k$ columns of $\mathbf{U}$, and $\mathbf{\Lambda}_k = \mathrm{diag}(\lambda_1, \dots, \lambda_k)$. The whitened summary is then
\begin{equation}
\hat{t} \;=\; \mathbf{W} \left( t - \bar{t} \right),
\qquad
\mathbf{W} \;=\; \mathbf{\Lambda}_k^{-1/2}\, \mathbf{U}_k^{\mathrm{T}}\, \mathbf{D}^{-1},
    \label{eq:whitening}
\end{equation}
so that $\hat{t}$ has dimension $k$, zero mean and unit covariance by construction. Note that before truncation \Cref{eq:whitening} is an invertible linear map and therefore preserves the information content in $t$ \citep{tegmark1997karhunen}. The NLE model is then trained on $p(\hat{t} \mid \theta)$ in place of $p(t \mid \theta)$.

In our approach, $\bar{t}$ and $\mathbf{W}$ are estimated once from the low-fidelity summaries and fixed thereafter. We apply the same fixed transformation to the high-fidelity summaries, so any systematic offset between the two fidelities is preserved and must be adapted to during fine-tuning. Note that \Cref{eq:whitening} removes only second-order structure, so all the residual non-Gaussianity and higher-order correlations must still be learned for the NLE model. We discuss our choice for the truncated dimensionality $k$ in \Cref{sec:NLE_results}.

\subsubsection{Model ensembles}
\label{sec:model_ensembles}

We explore a model ensembling scheme to improve performance under very limited dataset sizes. Calibration error is caused by modeling errors in the neural density estimation stage.  An appealingly simple approach to mitigate modeling errors in neural networks is to train multiple models and average their predictions \citep{dietterich2000ensemble,lakshminarayanan2017simple,alsing2019fast, hermans2022crisis, yao2024simulation}. This approach is inherently limited by any inductive biases in neural network architectures, which restrict the model ensemble to average over a subset of epistemic uncertainties. Nonetheless, for small dataset sizes this strategy can provide a practical and robust improvement by reducing variance in the learned posterior or likelihood estimates. For an ensemble of $M$ models $\{q_m(\cdot)\}_{m=1}^M$, we take a simple arithmetic mean of the estimated density.

To produce ensemble samples with NPE, we sample from each model independently and combine the samples. For NLE we use the mean estimate of the likelihood to perform MCMC sampling. We perform no ensemble weighting (see e.g. \citealp{loaiza2025deep} for a theoretical justification for this choice and its connection to Bayesian model averaging).

In this work, we construct ensembles by independently fine-tuning multiple instances of the pre-trained models, each initialised from the same weights but fine-tuned with different training and validation splits. This procedure is computationally inexpensive relative to generating additional high-fidelity simulations, and serves as a complementary approach to the multifidelity schemes above by stabilising inference in the low-data regime.

\subsubsection{High-fidelity dataset sampling}

Throughout this work we test model performance as a function of the high-fidelity dataset size. While in practice one would use the full high-fidelity suite, here we subsample this dataset to emulate limited-data regimes. For most experiments, we randomly subsample cosmologies to construct reduced datasets. For the ensemble learning training, which uses relatively small datasets, we instead use a simple stratified sampling scheme \citep[see e.g.,][]{lohr2021sampling}. This partitions the candidate cosmologies into strata defined by quantile bins along a few key parameters $\{\Omega_\mathrm{m}, \sigma_8, w\}$ and draws from each stratum in proportion to its population (proportional allocation), producing a representative and balanced sample. We find this yields a modest improvement in performance. The precise implementation can be found in the software (see Data Availability). 

\subsection{Model details}

All density modelling is performed using rational-quadratic neural spline flows \citep[NSFs;][]{durkan2019neural} implemented using the \texttt{sbi} library \citep{tejerosbi2022code}. This includes the multiple stages of VMIM optimisation in the hybrid learning approach (\cref{fig:hybrid_architecture})

Our hybrid learning approach requires training two compression networks to separately process the bandpower and convergence maps for a given simulation. We compress the bandpower observations using a feedforward network with Gaussian error linear unit (GELU) activation functions \citep{hendrycks2016gaussian}.

Once the bandpower compression network is trained, its weights are frozen and its outputs concatenated with a convolutional neural network (CNN)-based field level compression model. During training, the same CNN is used to process both KiDS patches independently, with convergence maps for each of the 6 tomographic bins stacked channel-wise to produce observations of dimension (6, 1000, 100). The CNN is made up of several repeats of blocks consisting of residual convolutional layers and a downsampling layer, gradually reducing the spatial resolution of the inputs while increasing the channel depth (i.e. number of features). The final layer of the CNN performs spatial pooling of the final feature maps \citep[inspired by the mean pooling used in][]{villaescusa2022camels}. The resulting embeddings from each patch are then concatenated and passed through a shallow feedforward network to produce the final field-level compressed statistic. 

We use this weight-sharing CNN design since the underlying physics across the patches is the same (and the only difference in systematics are small differences in the shear biases, see \cref{sec:shape_noise_annd_bias}). This approach should therefore improve the performance of the CNN by increasing the effective dataset size it sees during training. To allow the network to account for the minor differences between the patches induced by different footprint geometries and shear biases, we augment the CNN with Feature-wise Linear Modulation layers (FiLM, \citealp{perez2018film}, widely used to include e.g. float or class conditioning information in CNNs; \citealp{brock2018large, dhariwal2021diffusion}). 

\subsection{Data preprocessing}

All mock observations are scaled before being input into the compression networks. The bandpowers are log-scaled and then standardised to have zero mean and unit variance. The convergence maps are also standard scaled (with the same scaling applied across all tomographic bins and patches for simplicity). The scaling transforms are computed and applied separately for the two different fidelity datasets. The convergence patches are zero-padded to a standardised dimension of (1000, 100) pixels. 

We also utilise a strong augmentation scheme on the convergence maps. While we do not add any extra noise to either the bandpowers or the maps, we perform random vertical flips, horizontal flips, and $180$ degrees rotations to each patch independently during training. This encourages the CNN to learn equivariance under these symmetries in the convergence maps.

The cosmological parameters are all min-max scaled between $[0,1]$ within the Gower Street prior. Since the limits of each 1-D prior are matched to Gower Street for the \texttt{GLASS} log-normal simulations, the meaning of each parameter in the scaled space is kept consistent across the two fidelities, likely making transfer learning easier.   

\subsection{Evaluation}

The primary goal of this study is to explore to what extent transfer learning can reduce the number of high-fidelity simulations required for accurate and precise inference. To evaluate these criteria we perform training with and without our multifidelity approach, varying the number of $N$-body large box simulations. We use $N$ to denote the number of simulations used for training and model selection. For all experiments, we report metrics on $60$ held-out test simulations (10\% of the Gower Street simulations) that were never seen during training. Note that, due to the large number of augmentations for a given $N$-body simulation box, the number of mock observations in dataset $\mathcal{D}$ is $\mathcal{D} = 80 \times N$.

Following prior work, we use three headline metrics to probe model informativeness, accuracy, and precision \citep{lueckmann2021benchmarking,delaunoy2022towards,saoulis2025transfer}. These are all complementary and any one metric is insufficient to completely describe the usefulness of a given model. 

We use the test loss given in \cref{eq:MI_lower_bound} for the NPE models to describe how well each model extracts cosmological information from the observations. We refer to this quantity as $\Delta$MI since it is equal to the mutual information up to an additive constant. Differences in $\Delta$MI between models are therefore differences the amount of information it retains from an observation.

To assess accuracy (i.e. posterior reliability), we use a calibration error metric, defined by comparing the empirical coverage of posterior credibility intervals to their nominal values across the test set. Specifically, for a set of $K$ credibility bins, we compute the observed frequency $\hat{p}_i$ with which the true parameters fall within each bin, and compare this to the expected frequency $p_i = 1/K$. The calibration error $\mathcal{C}$ is then defined as the relative mean squared error between these quantities:
\begin{equation}
\mathcal{C} = \frac{1}{K}\sum_{i=1}^{K} \left(\frac{\hat{p}_i - p_i}{p_i} \right)^2.
\end{equation}
This metric quantifies deviations from ideal calibration, and is sensitive to systematic biases and misestimated posterior uncertainties. The empirical coverage is estimated using  Tests of Accuracy with Random Points \citep[TARP,][]{lemos2023sampling}. We fix $K=100$. We bootstrap the estimated credibility level statistics produced by TARP 25 times to produce mean and standard error estimates of the credibility levels $\hat{p}_i$. To visualise the coverage statistics, we plot both the cumulative observed credibility interval distribution \textit{and} the ratio or ``excess coverage'' $\hat{p}_i/p_i$. We find this latter quantity is much more discriminative for identifying very minor calibration issues. 

In practice, we found that performing coverage testing over the full 9-dimensional posteriors explored here could conceal calibration issues in specific parameters. We therefore also ran coverage tests over smaller subsets and individual parameters. For the purposes of this work, we  report calibration performance on the key 3-D subset $\{\Omega_\mathrm{m}, \sigma_8, w\}$ in cases where this was necessary to identify significant calibration issues. For posterior constraint plots, we visualise this 3-D subset together with the derived parameter $S_8 \equiv \sigma_8 \sqrt{\Omega_\mathrm{m}/0.3}$, consistent with standard weak lensing analyses. 

Finally, to assess precision, we compute a prior-referenced figure of merit (FoM) from the posterior covariance matrix $\mathbf{C}_{\mathrm{post}}$,
\begin{equation}
\mathrm{FoM} = \left( \frac{\det\!\left[\mathbf{C}_{\mathrm{prior}}\right]}{\det\!\left[\mathbf{C}_{\mathrm{post}}\right]} \right)^{1/2}.
\end{equation}
Both $\mathbf{C}_{\mathrm{prior}}$ and $\mathbf{C}_{\mathrm{post}}$ are estimated from samples, with the posterior covariances averaged over the entire test set.  We report this quantity for the full cosmological parameter set, as well as for selected lower-dimensional parameter combinations.

In addition, we report a per-parameter constraint shrinkage factor. For each parameter $\theta_i=\alpha$, letting $\sigma_{\alpha,\mathrm{prior}}$ and $\sigma_{\alpha,\mathrm{post}}$ denote the prior and posterior marginal standard deviations, we define
\begin{equation}
\mathcal{S}(\alpha) = \frac{\sigma_{\alpha,\mathrm{prior}}}{\sigma_{\alpha,\mathrm{post}}}.
\label{eq:shrinkage}
\end{equation}
Values $\mathcal{S}(\alpha)>1$ indicate that the posterior constraints are tighter than the prior for parameter $\alpha$.

\subsubsection{Prior specification}

NPE methods generally freeze-in the prior used during training. While this can be avoided, for instance through posterior transformation techniques \citep{greenberg2019automatic} or importance weighting during training, for simplicity we evaluate models against the Gower Street prior. This is modelled implicitly by the NPE methods, and explicitly for the NLE techniques. For the explicit model of the prior, we empirically model the joint prior over $\{\Omega_\mathrm{m}, \sigma_8\}$  in the Gower Street simulations with a neural spline flow, and use the analytic form of the Gower Street prior for all other parameters \citep[see][for the precise details]{jeffrey2025dark}.

We also test the ability of our NLE models to handle a realistic cosmological inference problem. For these results, we use a prior similar to the KiDS-Legacy main cosmic shear analysis \citep{wright2025kids}, which uses uniform priors on $S_8$ and the key cosmological parameters, as well as a joint Gaussian prior on the two intrinsic alignment parameters. Details are given in \cref{sec:kids_legacy_priors}.

\section{Results}
\label{sec:results}

\subsection{Architecture and hyperparameter optimisation}

We ran a number of experiments to improve the performance of both the neural compression algorithm and the density estimation neural networks. We performed a heuristic, manual optimisation of various architectures and design choices using both the \texttt{GLASS} log-normal mocks and the Gower Street simulations. We summarise the key architecture decisions here, and report the results of several ablation experiments to demonstrate the importance of these choices in \cref{tab:ablation_results}. The precise implementation of the architectures and hyperparameters can be found in the software (see Data Availability).

The two most important factors in training a useful map-level compression network were the choice of NDE and the hybrid learning scheme. We found that Masked Autoregressive Flows \citep[MAFs;][]{papamakarios2017masked} produced compression and density estimation with significantly worse MI performance than using neural spline flows (NSFs). We found that the hybrid learning scheme was also very helpful for reaching higher MI performance (and in significantly fewer epochs / training steps). We found that training the CNN directly on the $\kappa$ maps without hybrid learning took much longer (several days on an A100 GPU, as opposed to 16 hours with hybrid learning) to produce acceptable results, and was still not competitive with the hybrid approach. In addition, freezing the 2-point information during the map-level training period was essential, as otherwise the two-point compression network experienced significant overfitting that degraded the overall performance.

We found that using a feedforward network to compress the bandpower measurements marginally outperformed a shallow CNN (where each tomographic bin cross-spectrum was stacked channel-wise). Removing dropout across the network improved performance for all experiments. 

\begin{table}
\caption{
Ablation study for the transfer NPE approach evaluated on the test set using the full training dataset ($N=530$). 
Performance is reported in terms of mutual information ($\Delta$MI) and the 
$\Omega_\mathrm{m}$--$\sigma_8$ figure of merit (FoM). Higher values indicate better performance for both metrics.}
\label{tab:ablation_results}

\begin{tabular}{lcc}

\toprule

Ablation
& $\Delta$ Mutual Information
& FoM $\{\Omega_\mathrm{m}, \sigma_8 \}$  \\

\midrule

Our best configuration 
& 5.19 ± 0.06 & 24.98 ± 0.52 \\

\midrule

No hybrid learning
& 3.63 ± 0.13 & 6.24 ± 0.24 \\

MAF NDE head*
& 4.72 ± 0.04 & 26.59 ± 0.86  \\

Only mean pooling
& 4.90 ± 0.27 & 19.26 ± 5.02 \\
No cyclic LR schedule
& 4.90 ± 0.31 & 19.31 ± 4.91 \\
No patch ID conditioning
& 5.17 ± 0.02
& 23.89 ± 0.43 \\

\bottomrule
\end{tabular}
*The MAF models had poor calibration and therefore overstated the FoM.
\end{table}

 \begin{figure*}
    \includegraphics[width=\textwidth]{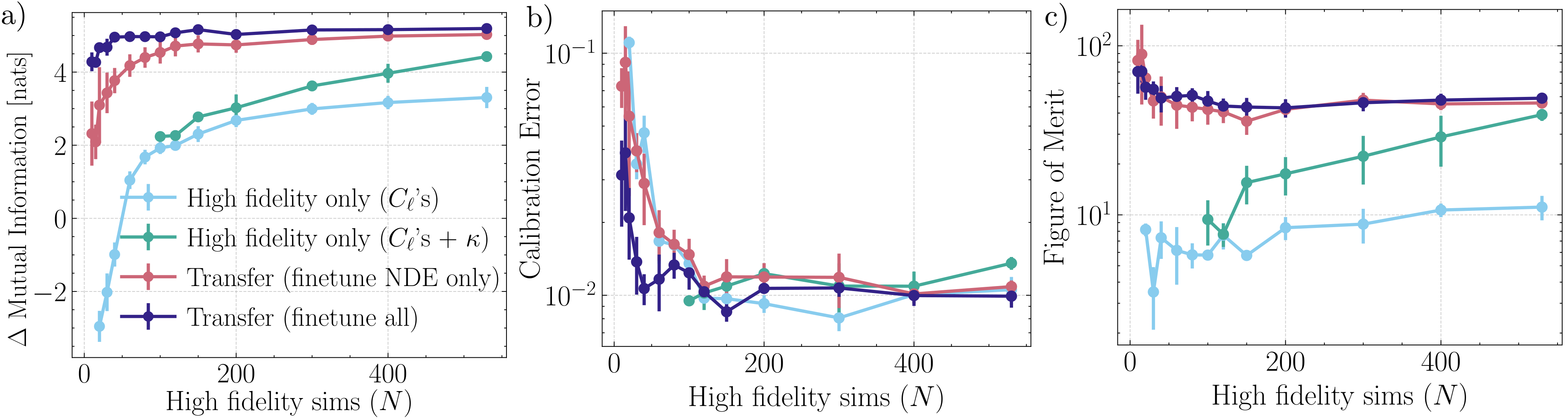}
    \caption{Inference results for the neural posterior estimation (NPE) approach on the full 9 parameter posterior distributions. The abscissa $N$ denotes the number of Gower Street large boxes used for training and model selection.  We present Gower Street only training (no pre-training), including and excluding higher-order information, against the results of our two NPE fine-tuning schemes. Panel a) shows the mutual information up to a fixed constant ($\Delta$MI) extracted by the compression and density estimator. Panel b) shows the posterior calibration error of each model. Panel c) shows the average figure of merit over the test set, which measures the relative constraining power of the posteriors with respect to the prior.}
    \label{fig:headline_metrics_9param}
\end{figure*} 
For the CNN compression model, we explored the effects of several architecture modifications. Providing patch label information to the CNN through the FiLM conditioning method provided very little benefit over no patch label, and better results than concatenating an extra channel to each $\kappa$ map image with the patch label. We adopted the mean spatial pooling operation at the end of the CNN from \citet{villaescusa2022camels}, and extended it to non-linearly combine the results of mean, maximum, and generalised mean pooling \citep[GeM;][]{radenovic2018fine}. This increases the expressivity of the final feature maps produced by the CNN, and led to slightly improved performance (\cref{tab:ablation_results}).

The results presented in this work used a CNN made up of blocks of two residual convolutional layers. Each block finishes by downsampling the spatial dimension by a factor of two. The blocks have a progressively increasing number of channels; we used 6 residual blocks with channels $[32, 32, 64, 64, 128, 256]$. The final feature map had a very small spatial extent, to which we applied the spatial pooling to produce patch-size agnostic feature maps. The resulting (non-linearly) combined aggregated features took dimension 256, which was then passed to a two-layer feedforward network to produce a final per-patch $\kappa$ map summary of dimension 256. We found that small changes to the number of residual blocks, convolutional channels, or embedding dimensions, did not produce significant changes in performance.

The two patch summaries were then concatenated and passed through a second shallow feedforward network. The feedforward network therefore took inputs of size $512$ and compressed them to produce the final beyond-two-point $t_\textrm{map}$ map-level summary. Although these CNN features should be broadly commensurate (i.e. mean similar things and are therefore comparable) across patches, we found that pooling across the two patch features (with the same mean, max, and generalised mean pooling as before) yielded worse results than a generic feedforward network. 

We set the summary dimensionality for the compression networks to be $\dim(t_\textrm{2-pt})=8$ and $\dim(t_\textrm{map}) =8$. This yields an embedding dimension of 16, which we passed through a linear layer to produce the final summary statistic $t$, also with $\dim (t) = 16$ (see \cref{fig:hybrid_architecture}). We found that increasing the compressed summary size did not improve performance.

For the neural density estimation head, we use a rational-quadratic neural spline flow (RQ-NSF) with 4 transformation blocks. Each block is parametrised by feedforward coupling networks with two layers that process the conditioning information. For the NPE models, we used feedforward layers with width 32 (finding larger sizes could worsen the stability of the compression training). For the NLE models, we found larger NDE models improved performance (perhaps since the inference dimension increases from $9$ to $16$), so used feedforward layers with width 64. 

We present the details on our training hyperparameters in Appendix \ref{sec:app_training_hyperparameters}. The most significant result was that the beyond-two-point compression model benefited from a cyclic learning rate schedule, which we report in \cref{tab:ablation_results}.

\subsection{Multifidelity NPE}
\subsubsection{High-fidelity only vs. transfer learning}

\label{sec:NPE_transfer_learning_results}
\begin{figure}
\includegraphics[width=\columnwidth]{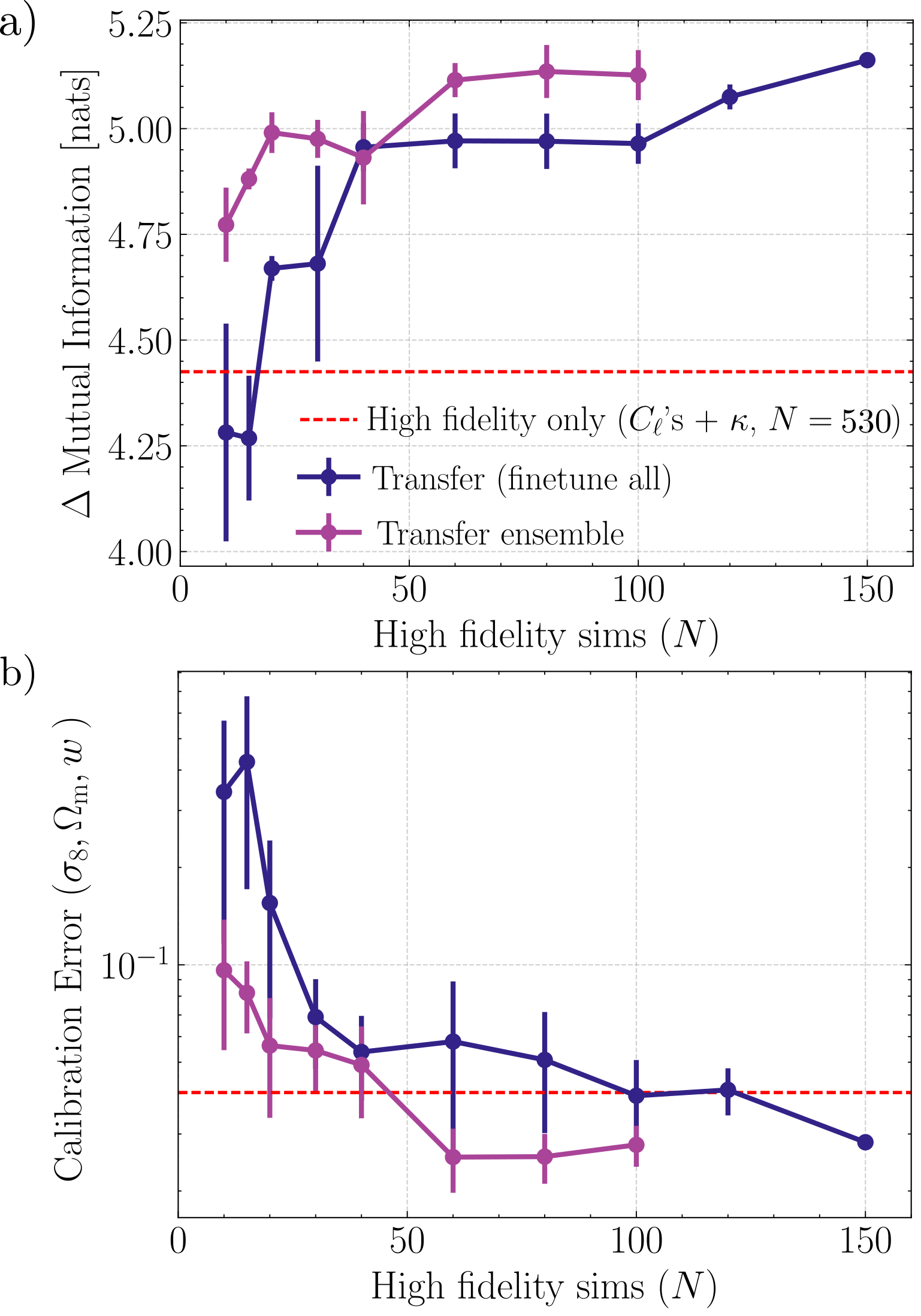}
    \caption{Inference results for the multifidelity neural posterior estimation (NPE) approaches on the full 9 parameter posterior distributions. A zoom-in of the results of the best performing Gower Street only model (trained with $N=530$ Gower Street simulations), compared with the  single-model transfer learning approach (``finetune all''), and the NPE ensembling approach. The ensemble consists of 9 models trained with the ``finetune all'' approach, each trained with a different train-validation split of the same $N$ cosmologies. Panel a) shows the mutual information up to an additive constant, $\Delta$MI, and panel b) shows the calibration error on the $\{\sigma_8, \Omega_\mathrm{m}, w\}$ subset of the modelled 9-D posteriors.}
    \label{fig:headline_metrics_NPE_ensemble}
\end{figure}

\begin{figure}
\includegraphics[width=\columnwidth]{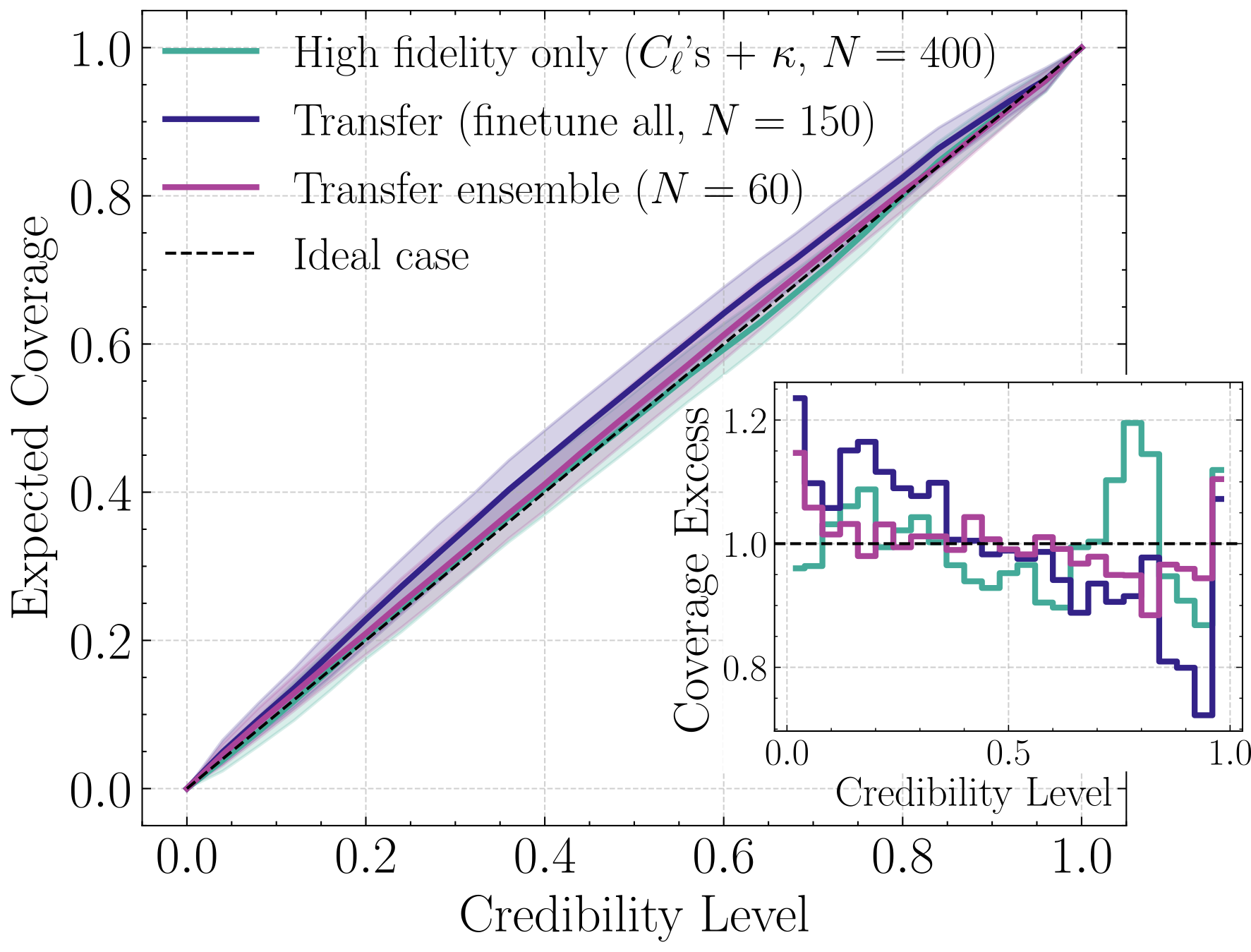}
    \caption{Calibration performance on the $\{\sigma_8, \Omega_\mathrm{m}, w\}$ subset of various representative NPE models on 190 held out Gower Street cosmologies (corresponding to $\sim3800$ mock observations). Credibility intervals are estimated using TARP, and $2\sigma$ uncertainties are estimated through bootstrapping.  The main panel shows the traditional (cumulative) observed credibility intervals, while the inset shows the ratio between observed vs. expectation of each credibility level  (i.e. its over-representation).  We compare each approach at various representative high-fidelity dataset sizes, denoted by $N$ in the legend. }
    \label{fig:coverage_NPE}
\end{figure}

 \begin{figure*}
    \includegraphics[width=\textwidth]{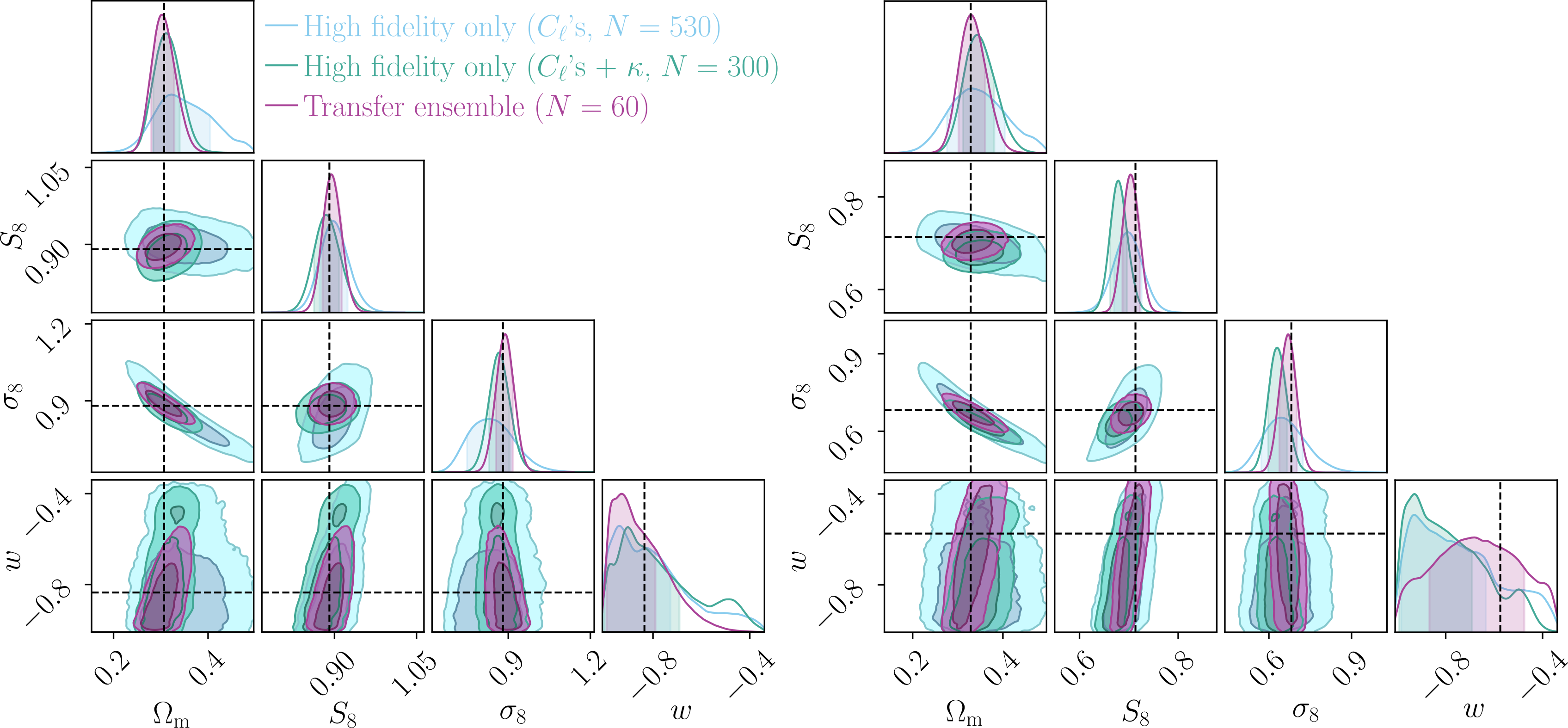}
    \caption{Two inference examples for a representative set of neural posterior estimation (NPE) approaches (left and right panels, respectively).  We present Gower Street only training (no pre-training), including and excluding higher-order information, against the results of the multifidelity transfer learning ensemble. $N$ denotes the number of Gower Street simulations used to train each model.}
    \label{fig:NPE_posterior_example}
\end{figure*}

We explore how model performance, with and without low-fidelity pre-training, performs as a function of the number of high-fidelity Gower Street simulations $N$. We select the same 60 Gower Street simulations to be held out as a fixed test set for all experiments. We repeat training (and pre-training) for each experiment three times to produce estimates of the metric mean and standard error, using different $N$-length subsets of the  non-test cosmologies for each repeat. 

\Cref{fig:headline_metrics_9param} shows the results from four different approaches as we vary the number of simulations available. The first two do not use the lower fidelity: a two-point statistic baseline (i.e. only the bandpower compression and inference in Stage I of \cref{fig:hybrid_architecture}), and the beyond-two-point statistic hybrid learning approach. These are compared with the multifidelity NPE transfer learning approaches, introduced as (i) and (ii) in \cref{sec:multidelity_approaches}.

Regarding the high-fidelity only models (that do not employ transfer learning and have their weights randomly initialised at the start of training), we find predictable results. Model performance --- measured across the MI, the calibration performance, and the FoM --- all gradually improve as a function of dataset size. In addition, the CNN begins to extract a large amount of beyond-two-point information from the observations after around $N=200$ Gower Street simulations, where the FoM (i.e. constraint tightness) is between a factor of $\times2-3$ greater than the bandpower-only model.

The calibration error starts very high for all high-fidelity-only models using the two-point information. This gradually saturates to $\mathcal{C} \approx 10^{-2}$, which we manually verified corresponds to near-perfect coverage for the full 9-dimensional posterior TARP tests. The remaining dominant contribution to the calibration errors is that our randomly selected test set of $60$ was a slightly unrepresentative sample across several parameters. We therefore conclude that these models reach acceptable calibration after $N\ge100$. 

The multifidelity approaches produce significantly better results. The compressor and density estimation head of these models were pre-trained with the hybrid learning scheme in \cref{fig:hybrid_architecture} on the low-fidelity log-normal mocks. The ``finetune NDE only'' approach freezes the compressor network and fine-tunes only the NDE head. The ``finetune all'' approach uses a low learning rate to fine-tune both the compression and NDE networks. We find that, even with very low dataset sizes $N\le100$, both transfer learning approaches achieve very high MI and FoM compared to the high-fidelity-only approaches. Both methods consistently achieve comparable calibration performance to the high-fidelity-only models after around $N\ge120$.

For the transfer learning models, the FoM is initially overly optimistic at low $N$ and decreases as more high-fidelity data are introduced, while the calibration error $\mathcal{C}$ remains elevated due to overconfidence. We find that the \texttt{GLASS} log-normal NPE models yield slightly tighter posterior contours than the well-calibrated models fine-tuned on the high-fidelity $N$-body simulations. This suggests that transfer learning initially inherits overly tight constraints from the lower-fidelity problem, and requires a moderate number of high-fidelity simulations ($N \sim 100$) for the NDE to ``unlearn'' this bias and recover reliable posterior widths.

We find that the ``finetune all'' strategy, in which the compression network is also fine-tuned, performs significantly better than fine-tuning only the NDE. Naïvely, one might assume that this is because the pre-trained compression is sub-optimal, and fine-tuning improves its ability to extract information from high-fidelity examples. However, as  $N\rightarrow530$ in \cref{fig:headline_metrics_9param}a), the frozen compression approach nearly recovers the MI and FoM of the ``finetune all'' method. These results indicate that the frozen compression network extracts almost as much information as when it is fine-tuned; in fact, it is the NDE that struggles to recover this information with smaller dataset sizes ($N<200$). We therefore hypothesise that finetuning the whole model improves performance by allowing parts of the compression network to absorb and adapt to the distribution shift, simplifying the transfer task for the (relatively shallow) NDE head. 

\Cref{fig:headline_metrics_9param} allows us to analyse the relative benefits of multifidelity NPE. We focus on the ``finetune all'' strategy, and only on models with acceptable calibration (dataset sizes $N\ge120$). We find that for limited dataset sizes ($N<400$), transfer learning provides a factor of between $\times2-3$ improvement in the constraining power --- comparable to the relative gain from including higher-order statistics. In addition, the transfer learning approach outperforms the high-fidelity only models even when using the entire dataset at $N=530$. The compression network comes with highly informative summaries, and it takes around $N\simeq100$ simulations to ensure the resulting posteriors are reliable and trustworthy. 

We test the effect of the dimensionality of the posterior inference problem in \cref{app:dimensionality_posterior_test}. We repeat the NPE approaches for a simplified 2-D inference problem, inferring just $\{\sigma_8, \Omega_\mathrm{m}\}$, and compare the results with the 9-D task. For the high-fidelity-only models, we find that the model performance is slightly degraded on the 9-D task (with lower FoM over $\{\sigma_8, \Omega_\mathrm{m}\}$ for larger dataset sizes $N>400$). On the other hand, the multifidelity approach performs near-identically in both cases, indicating that pre-training allows more flexibility and robustness independent of the difficulty of the inference task.

\subsubsection{NPE ensembles}

Our transfer learning approach produces promising finetuned models with very high MI and FoM metrics, even for small dataset sizes. However, the minor-to-moderate levels of calibration error at small dataset sizes shown in \Cref{fig:headline_metrics_9param} may prevent their application for analysis. As described in \cref{sec:model_ensembles}, ensembles provide a potential remedy to improve calibration in the very low data regime. 

We run several experiments ranging from $N\in[10, 100]$ training 9-member NPE model ensembles, and present the results in \cref{fig:headline_metrics_NPE_ensemble}. These are all fine-tuned from the same pre-trained inference network, with the only difference being the train-validation split of the $N$ cosmologies. We find that ensembling produces models with higher average MI, with significant benefits at the smallest dataset sizes. In fact, a fine-tuned model ensemble with only $N=10$ Gower Street simulations extracts more information from the observations than models trained from scratch with $N=530$ simulations (though, as \cref{fig:headline_metrics_NPE_ensemble}b) shows, such models are not yet calibrated). This represents over a $\times50$ reduction in the number simulations required to produce highly informative, field-level compression schemes.

Most importantly, model ensembles systematically reduce the calibration error. \Cref{fig:headline_metrics_NPE_ensemble} shows the calibration error on the key subset of parameters $\{\sigma_8, \Omega_\mathrm{m}, w\}$. This shows that the model ensemble approach achieves acceptable calibration with only $N=60$ Gower Street simulations.

\Cref{fig:coverage_NPE} shows the coverage plots for several representative models: high-fidelity only training, single-model transfer learning and model ensemble transfer learning. We compute the credibility interval statistics over $190$ cosmologies not seen during training (the test set combined with 130 unused simulations), since we found empirically that the held out test set of $60$ simulations was a slightly biased and unrepresentative sample. All three models produce acceptable coverage statistics. The ensemble transfer learning approach, with $N=60$ simulations, and the high-fidelity only approach with $N=400$, both produce near-perfect coverage. The single-model transfer learning approach yields very minor biases: at $N=150$, for instance, has bias on the order of $\sim0.2\sigma$. This level of bias has been deemed acceptable in previous SBI studies in cosmology \citep{prat2026dark}.

\Cref{fig:NPE_posterior_example} shows two representative examples of inference using the multifidelity ensemble approach, compared against high-fidelity only training with and without higher-order information. The transfer learning approach yields very tight constraints on $\{\sigma_8, \Omega_\mathrm{m}\}$ relative to the two-point analysis, and produces tighter constraints on both $\{\sigma_8, \Omega_\mathrm{m}\}$ than high-fidelity-only field-level models that use $\times5$-$10$ more simulations. It is also more capable of constraining $w$ relative to the high-fidelity-only methods. We explore this in more detail in \cref{sec:model_constraining_power}.

Note that all models in \cref{fig:NPE_posterior_example} are well-calibrated: the difference in constraining power between each model is therefore explained by the relative strengths of each compression algorithm. This reinforces the results shown in Figs.~\ref{fig:headline_metrics_9param} \& \ref{fig:headline_metrics_NPE_ensemble}: pre-training on low-fidelity data can enable more constraining inference than traditional single-fidelity SBI with a fraction ($N=60$ vs. $N=530$) of the simulations. This is due to the fact that deep-learning-based compression algorithms are extremely data hungry, and fail to extract optimal summaries with limited datasets (as noted in e.g. \citealp{bairagi2025many} and \citealp{park2025dimensionality}). 

We show an example of the full 9 parameter posterior distributions in \cref{fig:9param_posterior_example}. We find that all methods, including the multifidelity ensemble, do not significantly constrain the remaining cosmological parameters $ \{m_\nu,  h, n_\mathrm{s}, \Omega_\mathrm{b}\}$ and the intrinsic alignment parameters $\{A_\textrm{IA}, \beta_\textrm{IA}\}$ beyond their priors. Note that although the results in \cref{fig:headline_metrics_9param} indicate that after $N\ge100$ the high-fidelity only approaches yield acceptable calibration, \cref{fig:9param_posterior_example} shows that at $N=300$, the beyond-two-point high-fidelity only model yields very unstable (i.e. irregular and multimodal, thus likely unphysical) posteriors for the unconstrained parameters (this can also be seen for the high-fidelity-only models for $w$ in \cref{fig:NPE_posterior_example}).

\begin{figure}
\includegraphics[width=\columnwidth]{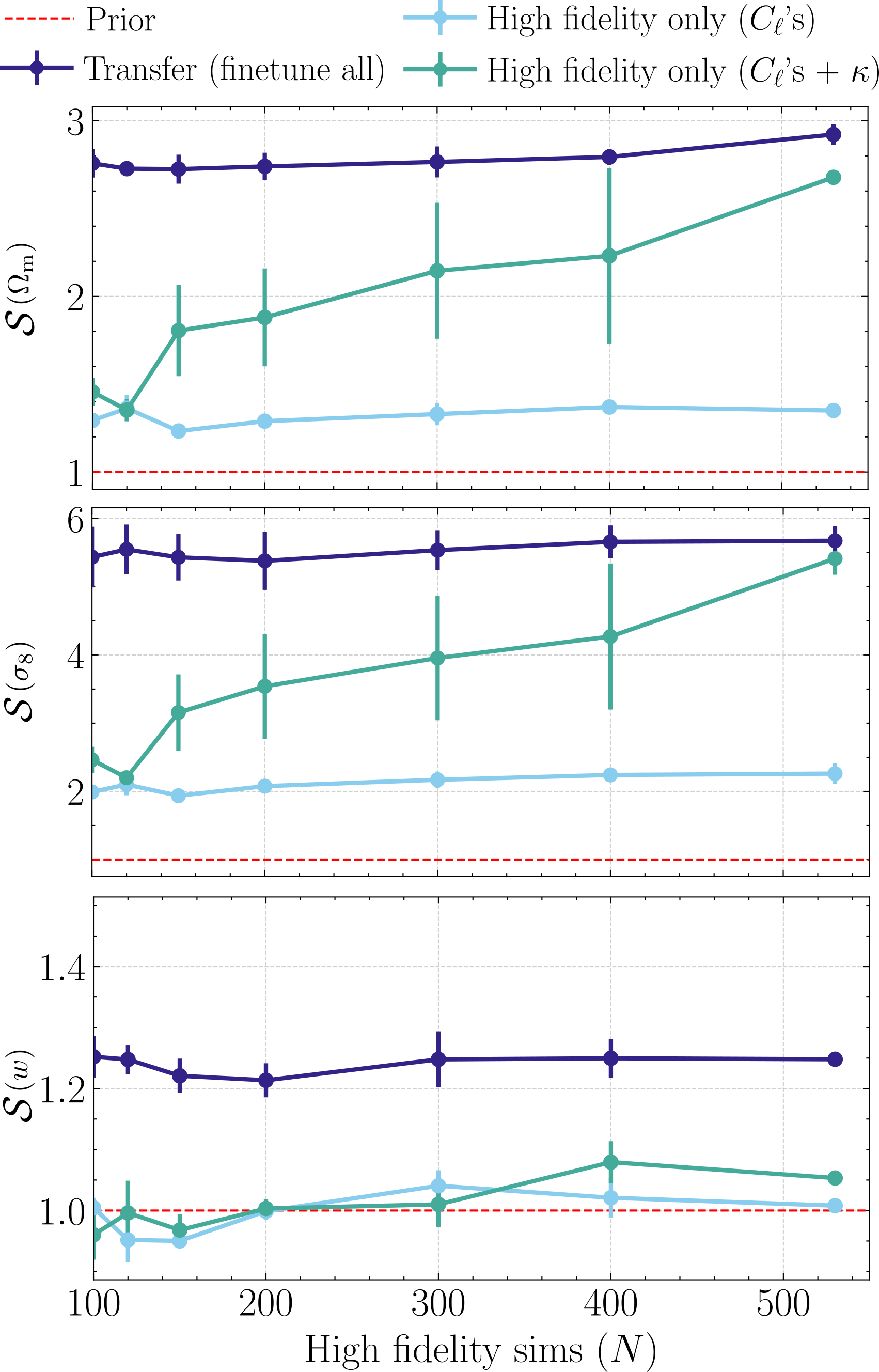}
    \caption{Posterior constraining power on the key cosmological parameters $\{\sigma_8, \Omega_\mathrm{m}, w\}$ as a function of high-fidelity training dataset size. The ordinate for each plot shows the relative per-parameter shrinkage of the posterior width $\sigma_\textrm{post}$ relative to the prior width $\sigma_\textrm{prior}$, $\mathcal{S}=\sigma_\textrm{prior}/\sigma_\textrm{post}$.}
    \label{fig:per_parameter_shrinkage}
\end{figure}
\subsubsection{Model constraining power}
\label{sec:model_constraining_power}

In order to better understand the constraining power of each approach, we probe the per dimension shrinkage $\mathcal{S}(\alpha)$ as defined in \cref{eq:shrinkage}. As shown in the posterior inference example in \cref{fig:9param_posterior_example}, only $\sigma_8$, $\Omega_\mathrm{m}$ and $w$ can be significantly constrained. We show the per-parameter shrinkage as a function of training dataset simulations $N$ in \cref{fig:per_parameter_shrinkage}.

The constraints on $\sigma_8$ and $\Omega_\mathrm{m}$ vary relatively predictably and similarly to the FoM shown in \cref{fig:headline_metrics_9param}; the transfer learning constraints are stable, while the high-fidelity-only approach gradually learns to tighten the constraints with more simulations. The multifidelity approach systematically outperforms single-fidelity training in-terms of constraints on $\sigma_8$ and $\Omega_\mathrm{m}$.

More surprising are the constraints on $w$ shown by $\mathcal{S}(w;N)$. As before, the pre-trained models are capable of producing more informative constraints than the prior. On the other hand, we find that the high-fidelity only model fails to constrain $w$ as well as the multifidelity approach even when using the full training suite of $N=530$ Gower Street simulations. This suggests that, for our dataset and ML-modelling approach, pre-training enables a qualitative shift in performance, yielding meaningful constraints on $w$ that are not recovered by the high-fidelity-only model even at the largest training set size considered. Note that our results in \cref{sec:NPE_transfer_learning_results} indicated that the frozen pre-trained compressor was nearly as informative as the fine-tuned one, implying that the compressor's sensitivity is mostly fixed after pre-training. The improvement therefore appears to be inherited from the analytic $w$ dependence encoded by \texttt{HMcode-2020} in the log-normal simulations, rather than from any additional $w$ information present in the $N$-body simulations alone.

We present a summary of posterior constraining power across all parameters individually in \cref{tab:constraining_powers}. We find that outside $\{\sigma_8, \Omega_\mathrm{m}, w\}$, the only parameter that can be constrained beyond the prior is the intrinsic alignment model halo mass exponent $\beta_\textrm{IA}$. All models display very little sensitivity to parameters $ \{m_\nu,  h, n_\mathrm{s}, \Omega_\mathrm{b}, A_\textrm{IA}\}$.

Interestingly, the beyond-two-point high-fidelity-only approach actually produces slightly tighter constraints on $\beta_\textrm{IA}$ than the multifidelity approach. While the difference is very minor, it is possible that this is an example of negative transfer, where pre-training ``freezes-in'' some modelled relationship (or perhaps insensitivity between) the observations and $\beta_\textrm{IA}$ that cannot be reversed during fine-tuning. It is possible (though at this stage speculation) that the halo-mass-dependent IA model produces more physically distinct features on $N$-body matter distributions (with real halos), as opposed to on the unphysical reproduction of halos in log-normal matter density fields.

\begin{table}
\caption{Posterior constraining power on the test set across each approach using the whole high-fidelity dataset ($N=530$). Per-dimension shrinkage $\mathcal{S}$, showing the relative tightness of the posterior over the prior, and figure of merit (FoM) computed over subsets of interest are shown; larger is better for both metrics. \textbf{Bold} values indicate significantly better performance than both alternatives.}
\label{tab:constraining_powers}

\begin{tabular}{lccc}

\toprule
& \multicolumn{2}{c}{High-fidelity-only} & Transfer \\
\cmidrule(lr){2-3} \cmidrule(lr){4-4}

Metric
& $C_\ell$'s
& $C_\ell$'s and $\kappa$ maps
& Finetune all \\

\midrule
\multicolumn{4}{l}{\textbf{Shrinkage}, $\mathcal{S}(\cdot)$} \\
\addlinespace
$ \Omega_\mathrm{m}$ & 1.35 ± 0.02 & 2.68 ± 0.03 & \textbf{2.92 ± 0.06} \\
$ \sigma_8$ & 2.26 ± 0.15 & 5.41 ± 0.24 & 5.67 ± 0.22 \\
$ w$ & 1.01 ± 0.01 & 1.05 ± 0.01 & \textbf{1.25 ± 0.01} \\
$ m_\nu$ & 1.01 ± 0.01 & 1.01 ± 0.01 & 1.00 ± 0.01 \\
$ h$ & 1.02 ± 0.04 & 1.01 ± 0.02 & 1.03 ± 0.01 \\
$ n_\mathrm{s}$ & 1.02 ± 0.01 & 1.00 ± 0.01 & 0.97 ± 0.01 \\
$ \Omega_\mathrm{b} h^2$ & 0.98 ± 0.02 & 1.01 ± 0.02 & 1.00 ± 0.01 \\
$ A_\textrm{IA}$ & 1.03 ± 0.00 & 1.00 ± 0.00 & 1.01 ± 0.01 \\
$ \beta_\textrm{IA}$ & 1.06 ± 0.01 & \textbf{1.11 ± 0.01} & 1.06 ± 0.01 \\
\addlinespace
\midrule
\multicolumn{4}{l}{\textbf{FoM}} \\
\addlinespace
$\Omega_\mathrm{m}, \sigma_8$ & 7.54 ± 0.69 & 21.89 ± 0.63 & \textbf{25.41 ± 1.21} \\
$\Omega_\mathrm{m}, w$ & 1.36 ± 0.02 & 3.38 ± 0.09 & \textbf{4.47 ± 0.14} \\
All parameters & 11.12 ± 1.84 & 39.15 ± 3.21 & \textbf{48.99 ± 3.11} \\
\bottomrule
\end{tabular}
\end{table}

Also shown in \cref{tab:constraining_powers} are figure of merits computed over subsets of the posterior. Over  $\{\sigma_8, \Omega_\mathrm{m}\}$, we find our multifidelity approach yields $\times3.5$ tighter constraints than the two-point only approach and around $15\%$ tighter constraints than the high-fidelity-only $C_\ell$'s and $\kappa$ maps model. Over  $\{\Omega_\mathrm{m}, w\}$, the multifidelity approach produces $\times3$ tighter constraints than high-fidelity-only $C_\ell$'s-only, and $30\%$ tighter constraints than the $C_\ell$'s and $\kappa$ maps model.

 \begin{figure}
    \includegraphics[width=\columnwidth]{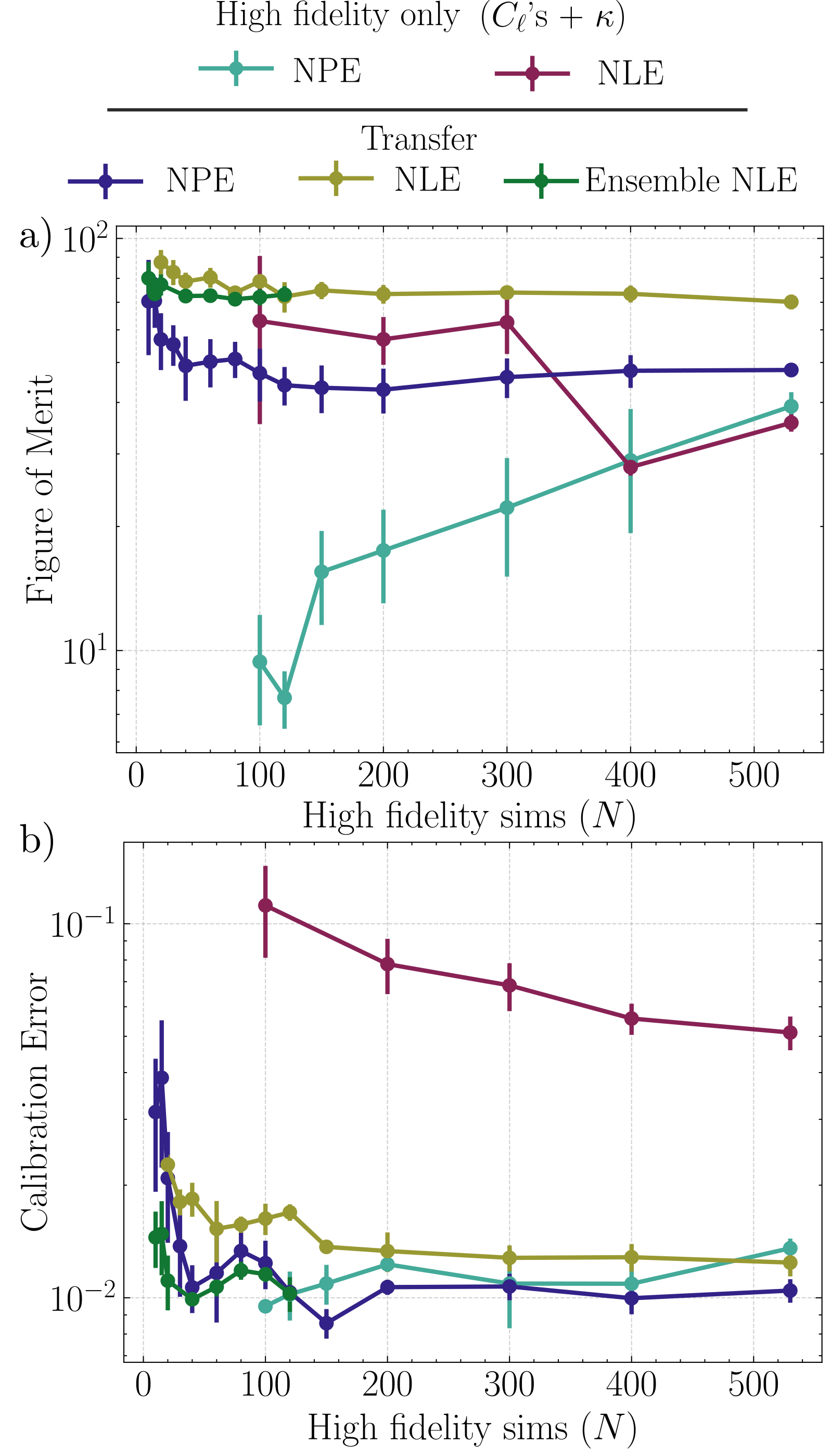}
    \caption{Inference results for the neural likelihood estimation (NLE) approach on the full 9 parameter posterior distributions, with the NPE approaches shown for comparison. The abscissa $N$ denotes the number of Gower Street large boxes available for training and model selection.  We present Gower Street only training (no pre-training) against the results of single-model and ensemble NLE fine-tuning schemes. Panel a) shows the average figure of merit over the test set, while panel b) shows the calibration error of each model.}
    \label{fig:NLE_headline_metrics}
\end{figure}

\begin{figure}
\includegraphics[width=\columnwidth]{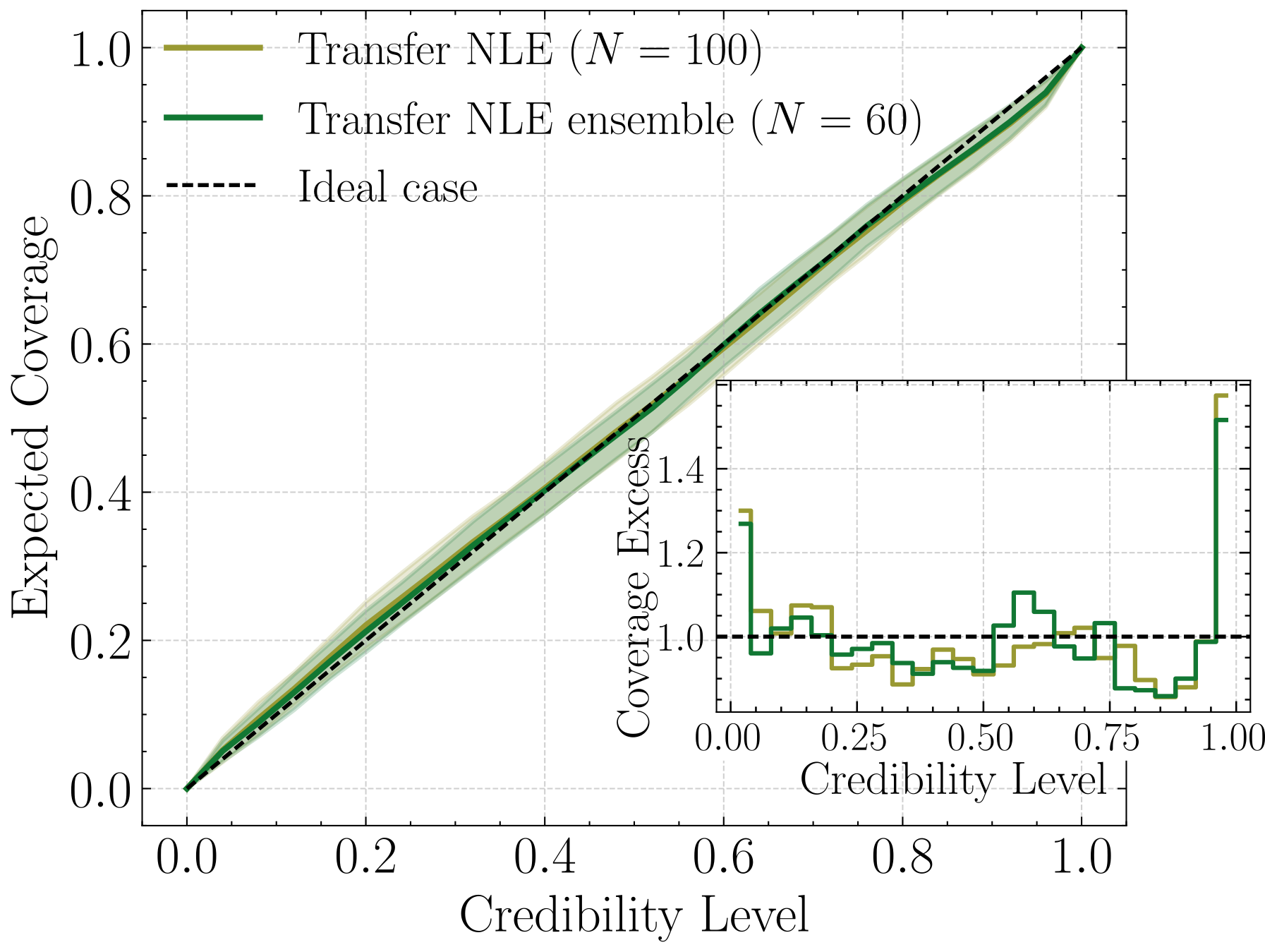}
    \caption{Calibration performance on the $\{\sigma_8, \Omega_\mathrm{m}, w\}$ subset of various representative NLE models on 190 held out Gower Street cosmologies. We use four augmentations for each cosmology, leading to $\sim760$ mock observations. Credibility intervals are estimated using TARP, and $2\sigma$ uncertainties are estimated through bootstrapping. The main panel shows the traditional (cumulative) observed credibility intervals, while the inset shows the ratio between observed vs. expectation of each credibility level  (i.e. its over-representation).  We show the single-model transfer learning approach with $N=100$ simulations (olive), and the transfer learning ensemble with $N=60$ simulations (dark green).}
    \label{fig:coverage_NLE}
\end{figure}

\subsection{Multifidelity NLE}
\label{sec:NLE_results}

 \begin{figure*}
    \includegraphics[width=\textwidth]{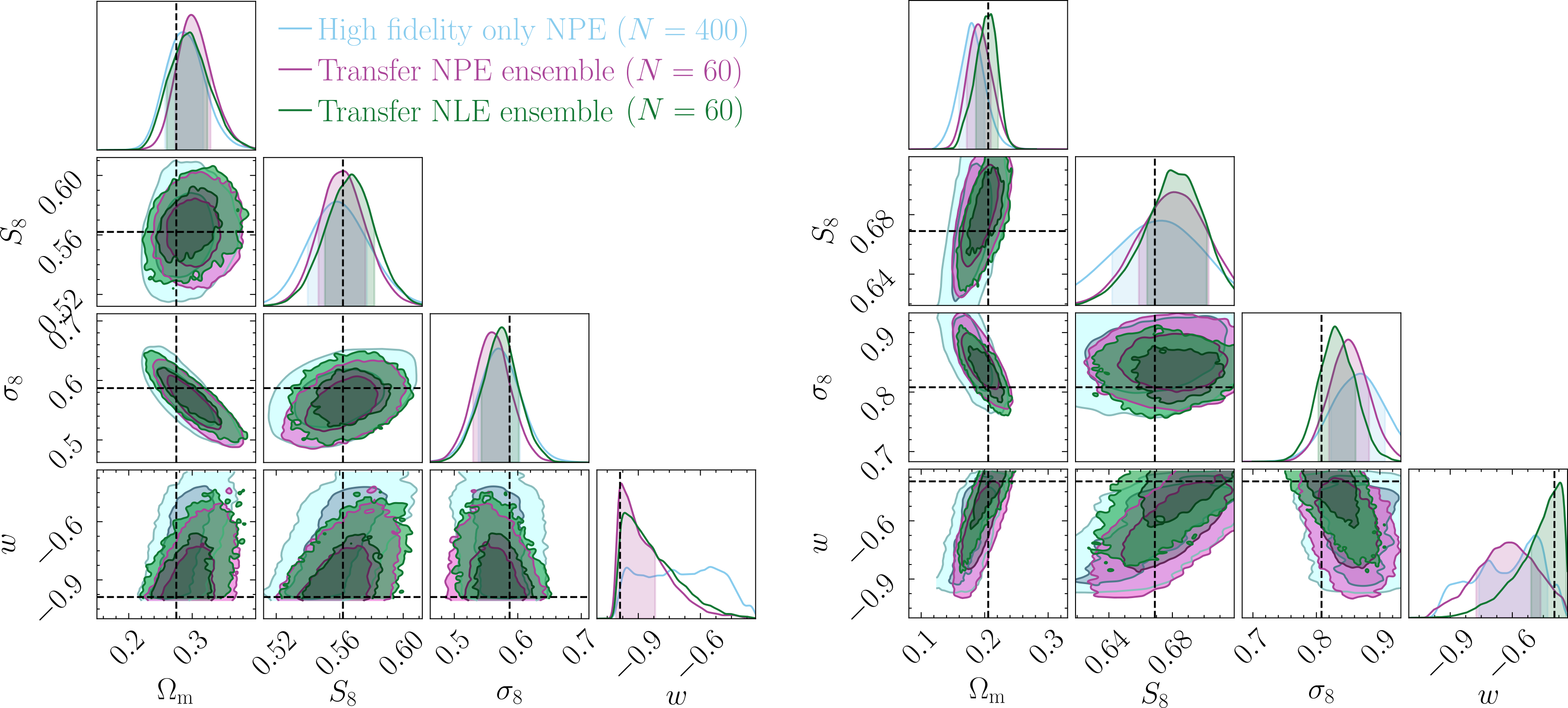}
    \caption{Two inference examples for the ensemble neural likelihood estimation (NLE) approach, compared with two NPE approaches. For visualisation purposes we show the $\{\sigma_8, \Omega_\mathrm{m}, w\}$ subset of the modelled 9 parameter posterior distributions, along with the derived parameter $S_8$. The right-hand example is a rare case where the different inference models yield slightly different posteriors: this is due to the combined effect of a different compression model and the slightly tighter constraints achieved by the NLE approach. }
    \label{fig:NLE_posterior_example}
\end{figure*}

We next explore multifidelity NLE, using the approach specified in \Cref{sec:multidelity_approaches} and \Cref{sec:summary_whitening}. We compare the results with transfer learning, whereby both the compressor and likelihood estimation NDE are pre-trained on \texttt{GLASS} log-normal mock observations, and without. We also test the benefits of using ensembles of NLE models. For the high-fidelity-only approach, we use the compressor trained using NPE with given simulation dataset size $N$ (i.e. the models shown in \Cref{fig:headline_metrics_9param})  to produce the training dataset $\{\theta, t\}$. For the multifidelity approaches, we use the compressors and density estimation models pre-trained on \texttt{GLASS} log-normal simulations directly, without any NPE fine-tuning on the higher-fidelity Gower Street simulations. For these multifidelity NLE models, we use the whitened and truncated summary $\hat{t}$ dimensionality of $k=8$ (described in \Cref{sec:summary_whitening}), finding that this comfortably preserves all the relevant information from the original $\dim(t)=16$ summaries. We explore the information content of the summaries as a function of the number of PCA components $k$ in \Cref{app:information_dimensions}.

We present the figure of merit and calibration error results for each approach in \cref{fig:NLE_headline_metrics}. We include two representative NPE experiments as baselines for comparison. We do not report the MI or the test loss; the former is more challenging to compute for NLE, and the latter is not commensurable across different summary statistics (and therefore across different trained neural compression models). For calibrated models, the NLE FoM should be similar to the NPE FoM (since both models rely on summaries with the same cosmological information).

\Cref{fig:NLE_headline_metrics} shows that the high-fidelity-only NLE models perform very poorly, even with the full high-fidelity Gower Street dataset. We find that its calibration error remains very high for all experiments, and is therefore not capable of providing reliable posterior estimates. The very high FoM at low dataset sizes $N$ (which returns to a more realistic FoM around $N>400$) of the Gower only NLE model is a result of severe overconfidence. 

These results are most likely caused by a combination of the small high-fidelity dataset and the relatively complex, high-dimensional raw summary space, which must be modelled probabilistically by the NLE model. We find that repeating high-fidelity-only NLE training for two-point compression models, which have $\dim (t)=8$, yields lower calibration error (\cref{app:NLE_bandpowers_metrics}). It is also possible that the relatively limited dataset leads to neural compression algorithms with poorly regularised latent spaces, leaving the NLE model a challenging task of modelling a highly irregular 16-dimensional likelihood. We provide some more discussion of this behaviour with reference to prior work in \cref{sec:discussion_existing_work}.

On the other hand, \Cref{fig:NLE_headline_metrics} shows that our multifidelity NLE approaches produce better posterior estimates with far fewer simulations. The calibration error is much lower than the high-fidelity-only approach, and comparable to both the single-model and ensemble performances of the NPE methods. We discuss the impact of the whitening procedure in \Cref{sec:app_whitening}, and demonstrate that the performance of NLE is significantly improved by the whitening and truncation procedure in \Cref{app:whitening_improvement}. Interestingly, the multifidelity NLE models achieve higher FoM than the multifidelity NPE approaches. Since this is not accompanied by a degradation in calibration, we interpret this as a genuine improvement in constraining power on cosmology enabled by the NLE approach. This is consistent with some prior findings in the literature \citep{lueckmann2021benchmarking}, though it may be partly enabled by our summary truncation procedure simplifying the density estimation task (see \Cref{app:information_dimensions}). 

Similar to the results in \Cref{fig:headline_metrics_NPE_ensemble}, we find that model ensembles produce systematically improved results over training just a single model of the likelihood. Model ensembling significantly reduces calibration error, especially for the smallest dataset sizes $N<100$. We present calibration curves for the multifidelity approaches in \cref{fig:coverage_NLE}, finding that both models are very well-calibrated. We also find a minor miscalibration issue: a very small fraction (around 2\%) of events are overrepresented in the first and last credibility level bins. This inadequately modelled tail is caused by a small number of cosmologies near the prior boundaries.

We present two representative examples of posterior inference using the ensemble NLE approach in \cref{fig:NLE_posterior_example}. We show the posteriors from two NPE approaches for reference. \Cref{fig:NLE_posterior_example} demonstrates that with $N=60$ high-fidelity simulations, the multifidelity NLE approach is capable of yielding accurate and constraining posteriors, comparable to and perhaps improving on those produced using NPE. 

\Cref{fig:NLE_posterior_example} also shows an example with minor disagreement between the posteriors produced by NPE and NLE. We intentionally select such an example to demonstrate two effects. The first is expected: the two posteriors used two different pre-trained compressors $F_\phi^*$, which produce slightly different (and therefore not sufficient) summaries with correspondingly different posteriors. We show an example of this effect in \cref{fig:repeated_training_example}. The second potential effect is the slightly improved posterior constraints produced by the NLE approach, as demonstrated by the FoM curve in \Cref{fig:NLE_headline_metrics}.

\subsection{KiDS-Legacy-like priors}
\label{sec:kids_legacy_priors}

The main advantage of NLE is that it can admit completely flexible priors during inference, which is a pre-requisite for standard cosmological analyses. As a proof-of-concept, we roughly replicate the KiDS-Legacy main analysis, with tophat priors on most parameters. The main differences are that we include $w$ as an extra inference parameter and do not fix the neutrino mass $m_\nu$ (though this latter choice likely has only minor impact on our final posteriors). 

The exact prior we use is shown in \cref{tab:priors}. Since the Gower Street simulation suite does not fully cover the KiDS-Legacy fiducial analysis priors, we truncate several parameters to ensure sampling never reaches regions where the NLE model will fail. 

\begin{table}
\centering
\caption{Prior range for our KiDS-Legacy-like mock analysis. Ranges $[$lower, upper$]$ denote the ranges of tophat priors. The values for the mean and covariances of the NLA-M model parameters are given in \citet{wright2025kids}, originally from \citet{fortuna2025kids}. The limits of the prior ranges are restricted to that of the Gower Street simulations.}
\begin{tabular}{cc}
\toprule
{Parameter} & {Prior Range} \\
\midrule
$S_8$ & [0.5, 0.9] \\
$\Omega_\mathrm{m}h^2$ & [0.074, 0.204] \\
$w$ & [$-1$, $-1/3$] \\
$\Omega_\mathrm{b}h^2$ & [0.022, 0.0228] \\
$n_\mathrm{s}$ & [0.948, 0.984] \\
$h$ & [0.64, 0.78] \\
$m_\nu$ & [0.06, 0.14] \\
$A_{\rm IA}, \beta_{\rm IA}$ & $\mathcal{N}(\mu_{A, \beta}, C_{A ,\beta})$ \\
\bottomrule
\end{tabular}
\vspace{0.5em}

\small
\label{tab:priors}
\end{table}

 \begin{figure*}
    \includegraphics[width=\textwidth]{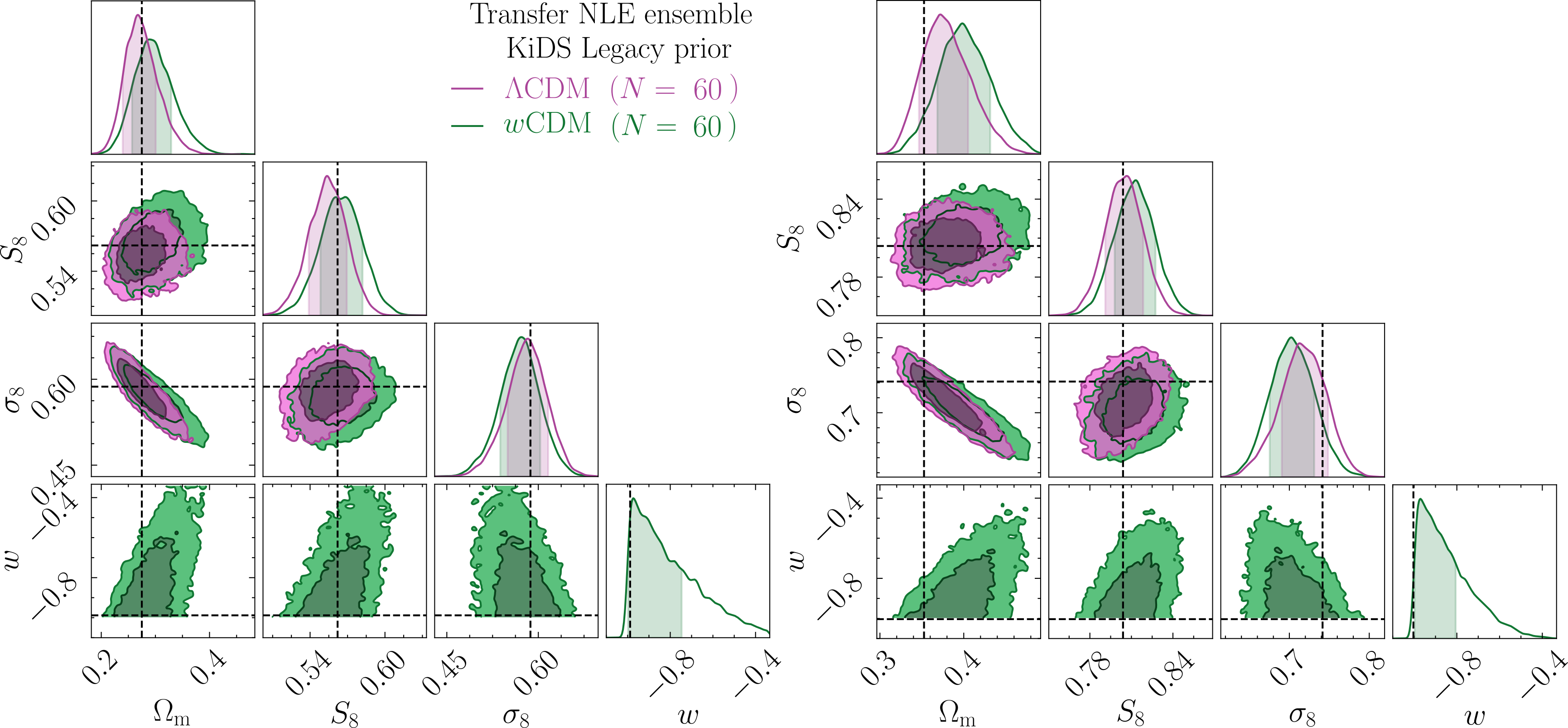}
    \caption{Two inference examples for the ensemble neural likelihood estimation (NLE) approach with KiDS-legacy-like  $w$CDM and $\Lambda$CDM priors (where $w$ is fixed to $-1$). For visualisation purposes we show the $\{\sigma_8, \Omega_\mathrm{m}, w\}$ subset of the modelled 9 parameter posterior distributions, along with the derived parameter $S_8$. $N$ denotes the number of Gower Street simulations used for training.  }
    \label{fig:LCDM_wCDM_example}
\end{figure*}

We use the NLE ensemble model that was used for \cref{fig:NLE_posterior_example}, trained with only $N=60$ Gower Street simulations, to perform inference on both $w$CDM and $\Lambda$CDM cosmologies. We present the results in \cref{fig:LCDM_wCDM_example} for two test cosmologies, with $S_8$ and $\Omega_\mathrm{m}$ similar to the estimates produced by the KiDS-Legacy cosmic shear analysis, and $w$ near $-1$. For the $\Lambda$CDM case $w$ is fixed to $-1$ in the NLE model. We find that, due to a mild positive degeneracy between $w$ and both $\Omega_\mathrm{m}$ and $S_8$, the $\Lambda$CDM cosmology prefers slightly lower values of $\Omega_\mathrm{m}$ and $S_8$ than $w$CDM. As expected, we also find slightly tighter constraints on these parameters in the $\Lambda$CDM parametrisation.

This test primarily serves as a proof-of-concept, demonstrating that our NLE models can be robustly adapted to custom priors as required for routine cosmological analyses. Nonetheless, we can provide some tentative comparison between the forecasts produced here and the constraints of traditional two-point cosmic shear analysis on the KiDS-Legacy data. For the $\Lambda$CDM case, for cosmologies with $S_8\in[0.75,0.85]$ and $w<-0.9$, we find average 68\% credible intervals of $\pm 0.023$ on $\Omega_\mathrm{m}$ and $\pm 0.015$ on $S_8$ over 50 mock observations, yielding similar constraints to KiDS-Legacy on $S_8$ and twice as constraining on $\Omega_\mathrm{m}$ \citep{wright2025kids, reischke2025kids}. For the $w$CDM case, \citet{reischke2025kids} produced posteriors with the KiDS-Legacy dataset with constraints $S_8 = 0.809^{+0.040}_{-0.041}$, $\Omega_\mathrm{m} = 0.332^{+0.058}_{-0.059}$, and $w = -1.10^{+0.45}_{-0.47}$. These were produced with a tophat prior on $w \sim \mathcal{U}[-3,0]$, so they are not perfectly comparable to our results (where the prior was restricted to the Gower Street prior). Nonetheless, over the same set of cosmologies we find average 68\% credible intervals of $\pm 0.030$ on $\Omega_\mathrm{m}$, $\pm 0.018$ on $S_8$, and $\pm 0.14$ on $w$. Even with the conservative assumption that our $w$ constraints would roughly double if $w<-1$ were considered in the prior, this corresponds to a near doubling of the tightness of the constraints across all key parameters $\Omega_\mathrm{m}$, $S_8$ and $w$.

We emphasise that these tests are intended primarily as a proof-of-concept demonstration that multifidelity SBI can be applied to a realistic cosmological analysis. The forecasted constraints presented here should therefore be interpreted with appropriate caution. A full analysis on real data will require validating the pipeline's robustness to unmodelled effects such as baryonic feedback, which could in turn motivate more conservative choices for the scale cuts or map smoothing.

\section{Discussion}
\label{sec:discussion}

\subsection{Advantages of multifidelity inference}

This work has demonstrated that transfer learning confers a wide range of benefits for enabling accurate and precise cosmological inference. We summarise our findings as follows:
\begin{enumerate}
    \item Highly informative compression schemes can be learned from lower fidelity simulations, and adapted to high-fidelity (i.e. at Stage III survey realism) observations with $\mathcal{O}(10)$ large box cosmological simulations. 
    \item Accurate inference, with precision and calibration similar to the tightest weak lensing constraints to-date, can be achieved with just $60$ simulations using transfer learning.
    \item Even for moderately large high-fidelity simulation datasets ($\sim500$ simulations), low-fidelity pre-training can produce tighter constraints and enable qualitatively new analysis; for instance, enabling significant constraints on $w$.
\end{enumerate}

Point (i) indicates that it is possible to construct highly informative beyond 2-point compression schemes that are very simulation-cheap. By comparison, traditional (i.e. not machine learning-based) compression techniques generally require many hundreds of simulations for computation of gradients and covariances \citep{alsing2019fast,park2025dimensionality}. This could aid the search for and improve our understanding of the effects of new physics beyond the two-point statistic level. For instance, comparisons between gravity-only and hydrodynamical simulations, or $\Lambda\textrm{CDM}$ against $w\textrm{CDM}$, could uncover distinct signatures at the field level. Moreover, it would be straightforward to apply such a search over different length-scales. These searches could be performed by probing the mutual information gain of different summary statistics \citep{piras2023robust,piras2024vae,piras2025lambda, prelogovic2024informative, sui2026evaluate}, or more directly by applying model interpretability techniques to the compression network \citep{gong2024c3nn,golshan2025massive,lahiry2025interpreting}

Point (ii) is most relevant for directly applying our approach to data. We have demonstrated that multifidelity SBI allows us to incorporate the most accurate physical models of cosmic structure formation directly during inference. We achieved a reduction in the number of $N$-body simulations required to perform accurate inference by a factor of $\sim10$. In the case of NLE, we also showed that standard high-fidelity-only inference failed for our task, and that our multifidelity approach actually made possible well-calibrated inference (given our simulation budget and summary statistic dimension). This is a promising result and could allow more accurate (or more diverse, beyond $\Lambda \textrm{CDM}$) physics to be included for a given simulation budget. In addition, different systematic effects could be evaluated cheaply. For instance, alternative galaxy biasing models or intrinsic alignment models, such as halo-based IA and tidal torquing, could be swapped in during the fine-tuning stage with relatively low computational overhead. 

Finally, point (iii) is an exciting prospect. For the Gower Street simulation suite and our mock observation pipeline, low-fidelity pre-training enabled meaningful constraints on $w$ that were not attainable using the available high-fidelity simulations alone. More broadly, pre-training increases flexibility in the design of summary statistics, allowing more expressive and informative summaries to be used with NLE, without being limited by the size of the high-fidelity dataset.

\subsection{Comparison to existing work}
\label{sec:discussion_existing_work}

Regarding the behaviour of transfer learning in multifidelity SBI, our findings are consistent with \citet{saoulis2025transfer}, who also demonstrate that highly informative neural compression models can be effectively transferred from low- to high-fidelity regimes using only a small number of high-fidelity simulations. Across both studies, the primary limitation in reducing the high-fidelity simulation budget is not compression, but rather the calibration of the neural density estimator during fine-tuning. In other words, once a suitable compression has been learned, multifidelity inference performance is largely governed by how well the density estimator adapts to the target high-fidelity distribution.

The most directly comparable prior work are DES Y3 results that have performed inference beyond the two-point level using the Gower Street simulations. \citet{jeffrey2025dark} trained a large ensemble of CNNs to process the DES Y3 convergence maps to estimate each cosmological parameter individually using a mean squared error loss. We expect that, due to the larger footprint area and greater number of galaxy measurements relative to KiDS-Legacy, DES Y3 maps should contain more cosmological information. 

\citet{jeffrey2025dark} achieve slightly tighter constraints on $\Omega_\mathrm{m}$ and $w$, and comparable constraints on $S_8$, relative to the synthetic forecasts in this work. The strong performance of our approach (relative to the expected constraints from each dataset) may be due to the pre-training and hybrid-learning compression scheme. However, it is also possible that different systematic models play a role --- \citet{jeffrey2025dark} marginalise over an uninformative prior over NLA parameters, for example. 

It is useful to place our high-fidelity-only results in the context of prior single-fidelity neural compression. \citet{bairagi2025many} find that the information a neural network extracts from the non-linear matter power spectrum only begins to saturate near $4000$ simulations, while \citet{park2025dimensionality} benchmark neural against linear compression at a budget of $\sim10^4$ simulations and find that their linear methods consistently outperform the neural compression techniques. Since both these works compress reduced-dimension summary statistics, they are already somewhat optimistic relative to our field-level approach. Prior field-level studies produce findings consistent with this work: \citet{bairagi2025patchnet} report that a CNN trained on the 3-D dark matter density field of $1500$ Quijote simulations gives worse constraints than the power spectrum alone, which they attribute to insufficient training simulations. The high-fidelity-only compression in \citet{saoulis2025transfer} required $N=6400$ maps for acceptable results. The behaviour of our own high-fidelity-only models, trained using a maximum of $N=530$ cosmologies, is therefore unsurprising, and the order-of-magnitude reduction we obtain from pre-training is in line with the field-level literature. We stress that these remain indicative comparisons, since each work uses different observables and infers different cosmological parameters.

We found that the high-fidelity-only NLE models performed surprisingly poorly. This is in contrast to several recent comparable works that report good constraints and calibration using standard, single-fidelity NLE. We note two major differences with prior work: these generally either use significantly more training data, as in \citet{thomsen2024des}, or use smaller dimensional summaries ($\dim (t) \leq 8$), as in \citet{jeffrey2025dark, prat2026dark}. We therefore interpret our findings as resulting from the high-dimensional ($\dim (t)=16$), poorly conditioned summary statistic space proving challenging to model for the likelihood emulator, given the relatively small dataset. We find further evidence of this through our results in \cref{app:NLE_bandpowers_metrics} that show two-point only compression, with $\dim (t)=8$, performed better than the beyond-two-point approach. This issue may be exacerbated by the VMIM optimisation procedure, which is somewhat agnostic to how easily downstream models can access the information content in the learned summaries. The hybrid learning approach could be adapted to produce smaller dimensional summaries, for instance by further compressing the two-point and convergence map summaries jointly. Nonetheless, our multifidelity approach enables more flexibility in the summary statistic dimension and ensures accurate modelling of the likelihood even with very limited simulation datasets ($N\sim60$).

\subsection{Future directions}

\subsubsection{Improving multifidelity transfer learning}

We tested two NPE transfer learning approaches described in \cref{sec:multidelity_approaches}: one that froze the pre-trained compression network and fine-tuned only the NDE head, and another that fine-tuned both the compression network and the NDE. Fine-tuning the full model consistently performed better, particularly for small training sets, than freezing the compressor. However, this is somewhat undesirable in practice, since it substantially increases the computational cost of fine-tuning and typically requires GPU resources. It also complicates downstream analyses, as the compression algorithm and resulting summaries are adapted for each new task (and may therefore cease to be comparable). 

We demonstrated that the main difference in performance between the two NPE schemes was not compression quality but instead the inflexiblity of the NDE when fine-tuned on the new fidelity. This could be addressed in future work with shallow or linear adaptation layers in the compression network \citep{houlsby2019parameter, chen2020simple}, or fine-tuning only a small portion of the compression network.  This effect may have contributed to the slightly worse NLE performance when working with the raw summary statistics observed in this work, since the NLE approach used the pre-trained compression network that was frozen during fine-tuning (so resembles the first, ``finetune NDE only'' approach).

We found that a simple whitening and truncation step of the summaries was required to ensure well-calibrated NLE models with very few ($N\sim60$) high-fidelity simulations. Skipping this step led to small but persistent miscalibration in the standard NLE-based models, which manifested as minor overconfidence. Our pre-processing step fixed two major issues that made multifidelity training less sample efficient: the large dimensionality of the summary statistic, and a complex likelihood structure of the raw learned summaries $p(t\mid\theta)$. There are many possible avenues to further mitigate these issues: reducing summary dimensionality through a further non-linear compression stage; regularising the summary statistic during training with an extra loss term to simplify the structure of the likelihood; or more advanced domain adaptation techniques that could more directly impose the probabilistic structure of the high-fidelity task, for instance through distillation \citep{thiele2025simulation}.

Ensuring well-calibrated inference is essential for applying multifidelity SBI to cosmological observations. We found that the minimum usable high-fidelity dataset size was determined by NDE calibration. The natural question following from the results of this work is whether calibration could be guaranteed with even fewer simulations. This is an active area of research, and existing approaches may achieve this for instance by explicit pairing of simulations across fidelities to improve the high-fidelity modelling \citep[see e.g.][]{lee2024zooming, thiele2025simulation, hikida2025multilevel}. Also appealing are balanced or conservative posterior estimation techniques, which are designed to avoid overconfidence  \citep{delaunoy2022towards, delaunoy2024low}. One could also develop our ensembling approach to prioritise well-calibrated posteriors \citep{yao2024simulation}.

We did not investigate the impact of the size of the low-fidelity simulation suite used during pre-training. It is plausible that substantially larger pre-training datasets would yield correspondingly tighter posterior constraints. This expectation is consistent with the well-established scaling behaviour observed in ML systems, where downstream task performance typically improves as a power-law function of training set size and model capacity \citep{kaplan2020scaling, hoffmann2022training, grattafiori2024llama}. This line of enquiry is beginning to gain traction across the physical sciences \citep{frey2023neural,bairagi2025many, vigl2026neural}. 

\subsubsection{Addressing new or misspecified physics}

A practical challenge not addressed in this work is how to handle the introduction of different parameters between fidelities. This is of particular interest in multifidelity cosmology, where for instance baryonic feedback parameters (e.g. $T_\mathrm{AGN}$) or  different intrinsic alignment models (e.g. NLA vs. TATT) may need to be added or adapted after pre-training. Prior work has used a simple approach of including dummy dimensions during pre-training that can be activated at the fine-tuning stage \citep{lucas2025hybrid, krouglova2025multifidelity}. We flag this as a potential area for future work. 

Future work should also test how well the benefits of transfer learning demonstrated in this work will generalise across different pairs of simulation fidelities and observable choices. In our case, the gap between the log-normal fields used for pre-training and the high-fidelity $N$-body simulations used for fine-tuning is large, providing a stringent test of the approach. However, the effectiveness of transfer learning will depend on the domain gap between the fidelities. This consideration may be important in the context of next-generation surveys, where statistical uncertainties will be significantly reduced compared to Stage-III datasets such as KiDS. As observational noise decreases, discrepancies between simulation fidelities will become more directly imprinted in the data. In this regime, any mismatch between fidelities is likely to have a more pronounced impact on inference, potentially reducing the gains achievable through transfer learning. In this context, one could consider e.g. field-level correction and baryonification techniques to supplement the lower fidelity observations \citep{piras2023fast,zhou2025map, schneider2025baryonification, sharma2025field}.

Application of our methodology to observations will require testing the sensitivity of our approach to unmodelled physics and systematics, analogous to standard inference approaches. Comparable work has approached this by testing whether unmodelled physics, such as baryonic effects or differing levels of source clustering, lead to significant posterior bias \citep{jeffrey2025dark, prat2026dark, thomsen2024des}. These have found that ML-based compression algorithms appear relatively robust for Stage-III precision. Nonetheless, both detection of and ensuring robustness to model misspecification is important for improving both the interpretability of and community trust in these precision SBI techniques. 

Prior work has demonstrated that misspecification can be mitigated by identifying and removing summaries that are overly sensitive to systematic effects, and by aligning or recalibrating the simulated summaries toward the observed data summary \citep{huang2023learning, pierre2026mitigating}. In addition, domain adaptation techniques such as contrastive learning  have been used to make the compression algorithm robust to nuisance parameters \citep{roncoli2023domain, akhmetzhanova2024data, andrianomena2025towards}. While we used ensembles to reduce modelling errors, a well known misspecification detection technique is to use intra-ensemble-member disagreement as a proxy for epistemic uncertainty \citep{lakshminarayanan2017simple}. This has been used as an NDE convergence test (\citealp{alsing2019fast, von2025kids,jeffrey2025dark}, and more recently formalised by \citealp{alvey2026simulation} for explicit out-of-distribution testing), but future work should investigate the sensitivity of the deep neural compressors to misspecification through this deep ensemble approach. 

\section{Conclusion}
\label{sec:conclusions}

We have demonstrated that multifidelity simulation-based inference provides a practical route to field-level cosmological analysis under realistic computational constraints. By leveraging transfer learning, we showed that highly informative, beyond–two-point compression schemes can be learned from inexpensive simulations and adapted to high-fidelity data with only $\mathcal{O}(10)$ large-volume $N$-body simulations. We achieved accurate, well-calibrated inference with as few as $60$ high-fidelity simulations. We forecasted precision comparable to leading current cosmic shear analyses, though a real-data application is left to future work. Even when moderately large simulation sets were available, low-fidelity pre-training improved constraints and allowed a more flexible approach to inference. Together, these results highlight multifidelity SBI as a flexible and simulation-efficient framework for incorporating increasingly realistic physics into next-generation cosmological inference.

\section*{Acknowledgements}

We thank Joachim Harnois-Deraps for constructive comments on a draft of this manuscript. We are grateful to Joshua Williamson for helpful conversations in preparing for this work. 
AAS is supported by the STFC UCL Centre for Doctoral Training in Data Intensive Science (grant ST/W00674X/1) and by departmental and industry contributions. AAS was also supported by the A. G. Leventis Foundation educational grant scheme. DP was supported by a Swiss National Science Foundation (SNSF) grant, and by the SNF Sinergia grant CRSII5-193826 ``AstroSignals: A New Window on the Universe, with the New Generation of Large Radio-Astronomy Facilities''. ASM acknowledges funding from a Leverhulme Trust Research Leadership Award. AMGF is grateful to support from the UPFLOW project, funded by the European Research Council under the European Union's Horizon 2020 research and innovation program (grant agreement No 101001601). NJ and BJ acknowledge support by the ERC-selected UKRI Frontier Research Grant EP/Y03015X/1.

\section*{Data Availability}
\label{data_availability}

The \texttt{Python} software used to produce the results of this paper is available at \href{https://github.com/asaoulis/glass_gower_transfer}{https://github.com/asaoulis/glass\_gower\_transfer}. All the simulation data used are publicly available via the Gower Street simulations public release \href{http://www.star.ucl.ac.uk/GowerStreetSims/}{http://www.star.ucl.ac.uk/GowerStreetSims/}.

\bibliographystyle{mnras}
\bibliography{mnras_guide} 



\appendix
\section{Training hyperparameters}
\label{sec:app_training_hyperparameters}

We optimised several training hyperparameters to improve model performance. We found using smaller batch sizes significantly degraded performance, finding most models performed much better with e.g. a batch size of 128 compared to 32. We used a weight decay of 0.01 for all experiments \citep{krogh1991simple,loshchilov2017decoupled}, and performed gradient descent with an AdamW optimiser \citep{loshchilov2017decoupled}. We found that each stage of training benefited from slightly different training configurations, such as learning rates (LRs) and schedulers.

For the first stage of compression training, we trained the bandpower compression model using a batch size of 128 and a fixed learning rate of $4\times10^{-4}$. We trained the low-fidelity model for 80 epochs and the high-fidelity-only models for 40 epochs (for all dataset sizes). 

For the second stage of compression pre-training (training the CNN to compress convergence maps), we utilised a batch size of 100, which was roughly the memory limit of the hardware used for training (a single L40S or A100 GPU). For both fidelities, we used a cyclic learning rate scheduler, which ramps up and then decreases the learning rate from its maximum to $1/10$th of its maximum repeatedly over training (\citealp{smith2017cyclical}, shown to help with other neural compression models in cosmology, e.g. \citealp{villaescusa2022camels, saoulis2025transfer}). For the low (high) fidelity, we used a linear LR warmup of 2000 (500) steps and a baseline maximum LR of $2\times10^{-4}$ ($1\times10^{-4}$). We used a cyclic period of 6000 (2000) training steps. We found a cyclic learning rate schedule significantly sped up the convergence rate of the model. For some learning rates ($>4\times10^{-4}$) or schedulers (e.g. constant or exponential), the models would get stuck at a sub-optimal performance. 

For the low-fidelity, we performed CNN compression pre-training for around $65,000$ gradient descent steps ($\sim 80$ epochs) and saved the lowest validation model. Training convergence varied significantly between models even when using the optimised cyclic learning rate scheduler, indicating further efforts to improve training stability (either through compression architecture, NDE architecture, or training techniques) would be desirable. For the high-fidelity-only models, we performed several training runs with different dataset sizes $N$. We fixed the number of epochs to $100$ for all training runs, after which gains from training longer were very minor. 

For all pre-training and high-fidelity-only models, we used a train --- validation split of $90\%$ to $10\%$. 

For our NPE transfer learning results, we reduced the learning rate and used a constant learning rate schedule. For the ``finetune all'' experiments, which fine-tuned the compression networks as well as the density estimation head, we used a learning rate of $1\times10^{-5}$. For the ``finetune NDE only'' approach we used a constant learning rate of $1\times10^{-4}$. For $N<40$ experiments, we fine-tuned for 25 epochs, and reduced this to $10$ epochs for $N\ge40$ (finding fewer epochs were needed for convergence). Finding that for very small dataset sizes model selection was an important contributor to the overall performance, we used a train-validation split of $75\%$ to $25\%$. For the model ensembles, we used even more data for model selection: $67\%$ for training and $33\%$ for validation (rounded to the nearest integer fraction of the $N$ available simulations). 

For the NLE models, we trained the models for significantly longer, finding convergence took longer. We pre-trained the low-fidelity NLE NDE for 250 epochs with a learning rate of $1\times10^{-3}$. For fine-tuning, we trained each model for $50$ epochs and used a learning rate of $4\times10^{-4}$. We used a constant learning rate schedule. For the high-fidelity-only models, we encountered significant difficulties producing well-performing models.  Our best results were achieved by training each model for 300 epochs with an aggressive exponential decay learning rate schedule that varied the learning rate from $1\times10^{-3}$ to $1\times10^{-5}$.  We used a train-validation split of $80\%$ to $20\%$ for all experiments. 

\section{Whitening and truncation of the NLE summary space}
\label{sec:app_whitening}

We find that using the raw summary statistic space leads to slightly overconfident posteriors (underestimating the posterior uncertainties by around 5\%). This is heavily punished by our calibration error metric, leading to systematically higher miscalibration errors. This is in contrast to the truncated and whitened summary space NLE models.

As mentioned for the NPE approach in \cref{sec:NPE_transfer_learning_results}, the low-fidelity pre-training yielded tighter posteriors (and likelihoods) than in the high-fidelity case. The overconfidence of the raw summary NLE models is likely partly explained by the mismatch between fidelities, which gradually becomes corrected with more high-fidelity examples. However, unlike in the NPE and truncated NLE cases, the miscalibration does not fully disappear as the high-fidelity dataset size increases. 

\begin{figure}
\includegraphics[width=\columnwidth]{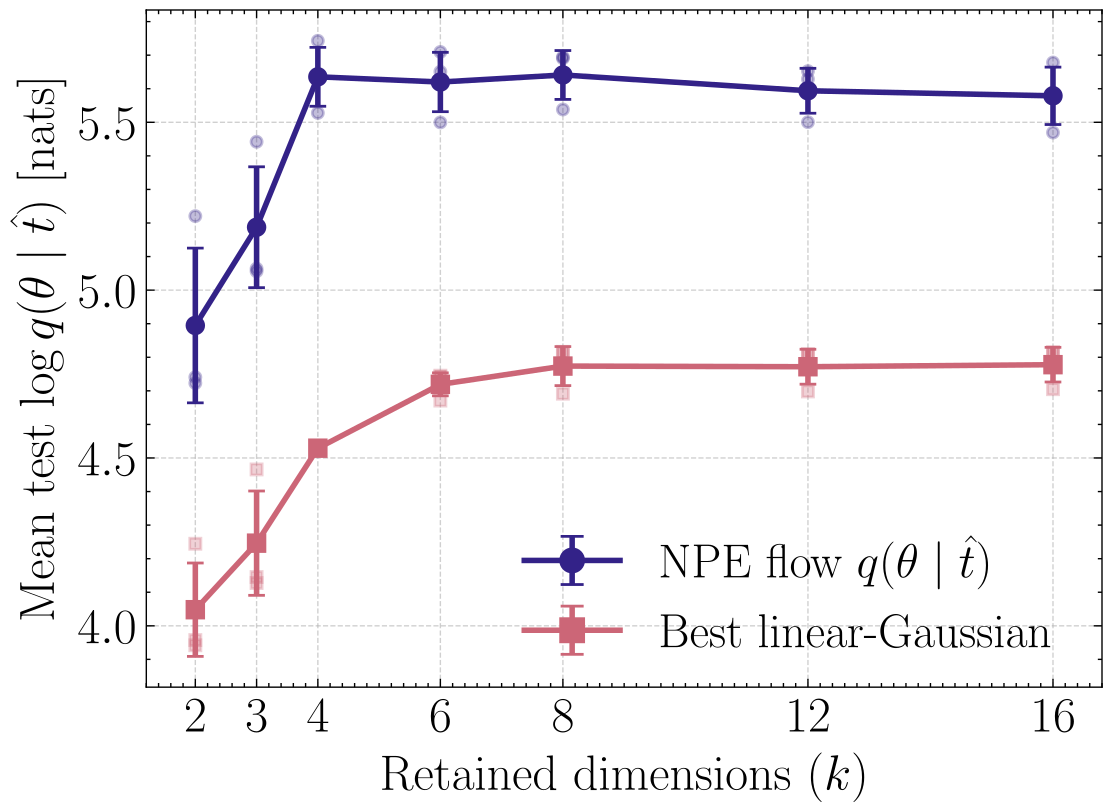}
    \caption{A sensitivity analysis into the PCA truncation dimension $k$ (i.e. the effect of keeping the $k$ leading eigenmodes of the summary statistic dataset). We explore both a linear Gaussian probe into the information content (assuming a Gaussian posterior), as well as re-training an NPE model to produce a non-linear information probe. We find that almost all the information relevant to cosmology is contained within the first $k=4$ eigenmodes.}
    \label{app:information_dimensions}
\end{figure}

Nonetheless, our truncation and whitening procedure enables accurate NLE inference, and requires fewer high-fidelity simulations to achieve this. We analysed the information content of the $k$ leading eigenmodes of the summary statistics $t$, with results shown in \Cref{app:information_dimensions}. We sweep across a range of truncation dimensions $k$, and fit a linear Gaussian model, as well as a fresh NPE neural spline flow to probe non-linear information content in the GLASS simulations.  We find that almost all the information relevant to the cosmological parameters $\theta$ is contained within the first $k=4$ eigenmodes. We use $k=8$ in the main text as a conservative choice to ensure negligible information loss from the original $\dim(t)=16$. Note \Cref{app:information_dimensions} suggests that the NPE models also (very slightly) degrade with higher summary dimension---though it is possible this is just noise. 

\begin{figure}
\includegraphics[width=\columnwidth]{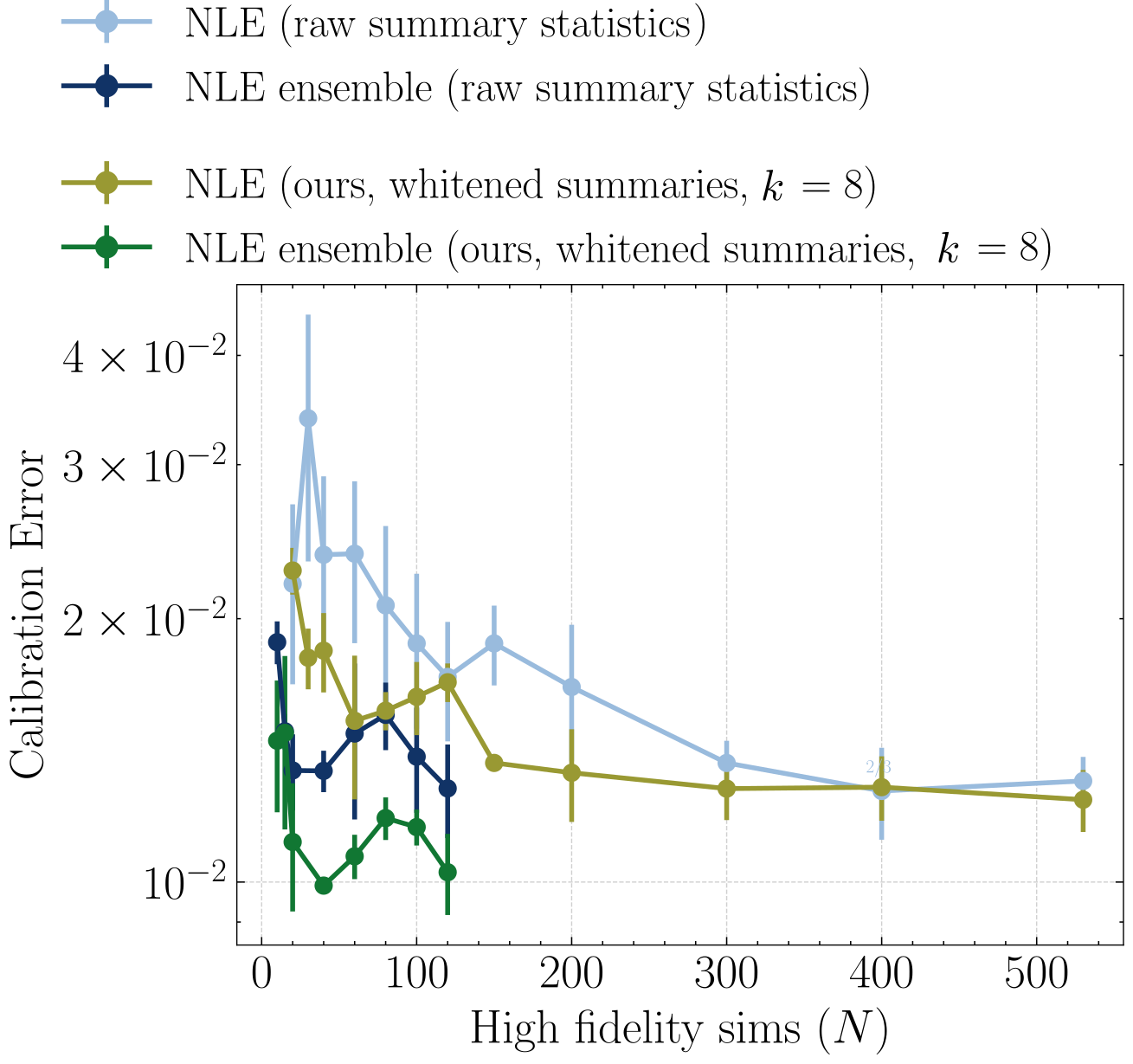}
    \caption{Calibration error comparison between the multifidelity NLE approach on the raw summary statistics (with $\dim(t)=16$) and the truncated, whitened summary statistics (with $\dim(\hat{t})=8$). For both the single model and ensemble approaches, our truncation and whitening substantially improves calibration, enabling accurate inference with fewer high-fidelity simulations.}
    \label{app:whitening_improvement}
\end{figure}

Finally, we test how our truncation and whitening improves performance as a function of high-fidelity simulations $N$ in \Cref{app:whitening_improvement}. We find that both the single model and ensemble NLE approaches significantly benefit from the procedure, producing better calibrated posteriors with fewer simulations. We verified that these improvements also translate to more accurate posterior constraints on the key cosmology subset $\{\Omega_\mathrm{m}, \sigma_8, w\}$. We conclude that reducing the dimensionality and complexity of the summary space through this simple linear transformation yields well-calibrated inference for NLE in far fewer simulations relative to the standard approach.

\section{Further examples and experiments}

\begin{figure}
\includegraphics[width=\columnwidth]{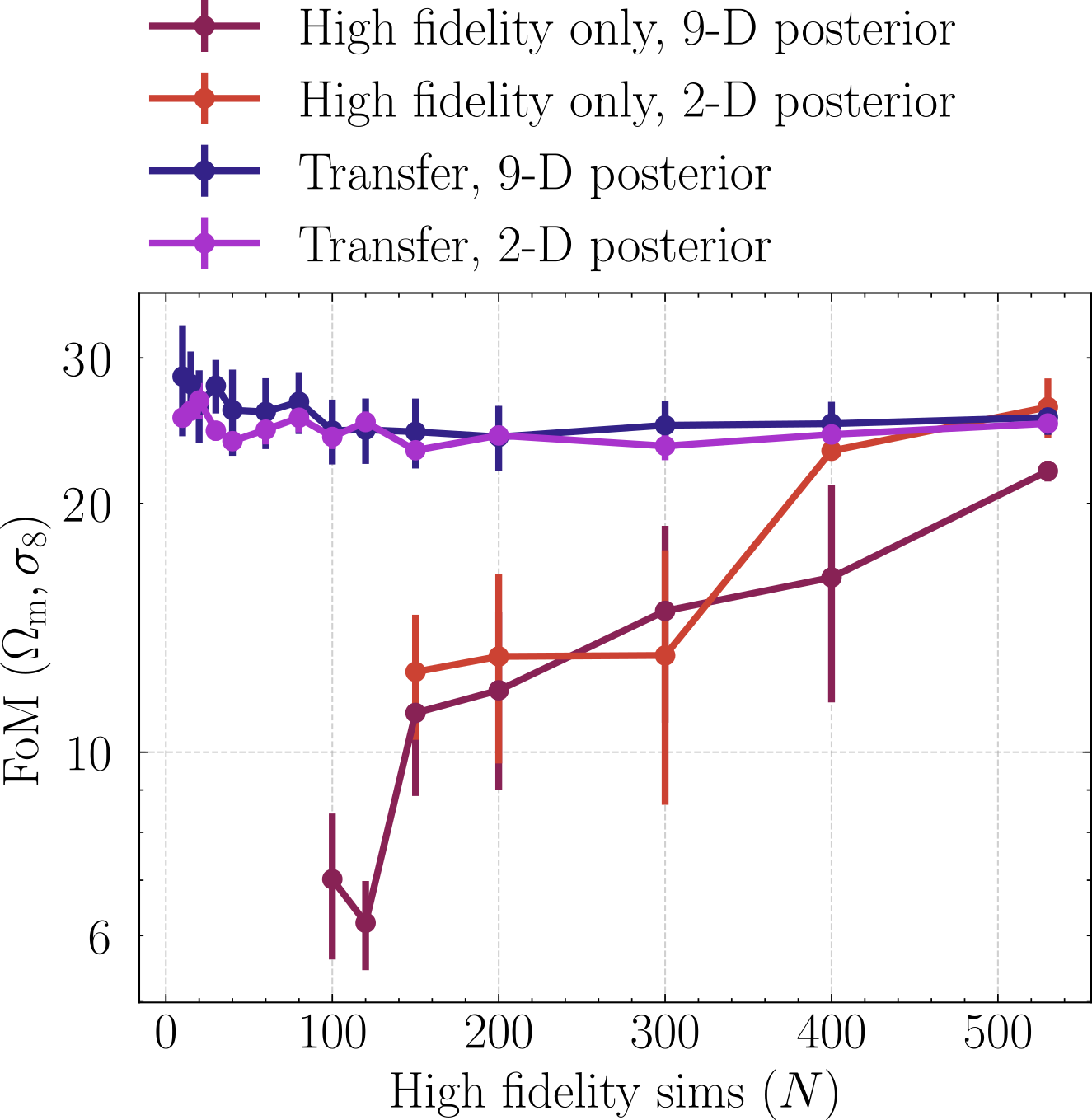}
    \caption{Figure of merit comparison on the modelled subset $\{\Omega_\mathrm{m}, \sigma_8\}$ between models learning the full 9-D posterior vs. those trained to only model the 2-D $\{\Omega_\mathrm{m}, \sigma_8\}$. All models are the beyond-two point hybrid learning models using pseudo-$C_\ell$'s and $\kappa$ maps as inputs. }
    \label{app:dimensionality_posterior_test}
\end{figure}

We present several extra figures to support and contextualise claims in the main text. 

\noindent
\cref{app:dimensionality_posterior_test} explores how the performance of the high-fidelity-only approach and the multifidelity approach varies for different posterior dimensionalities (2-D vs. 9-D). 

\begin{figure*}
    \includegraphics[width=\textwidth]{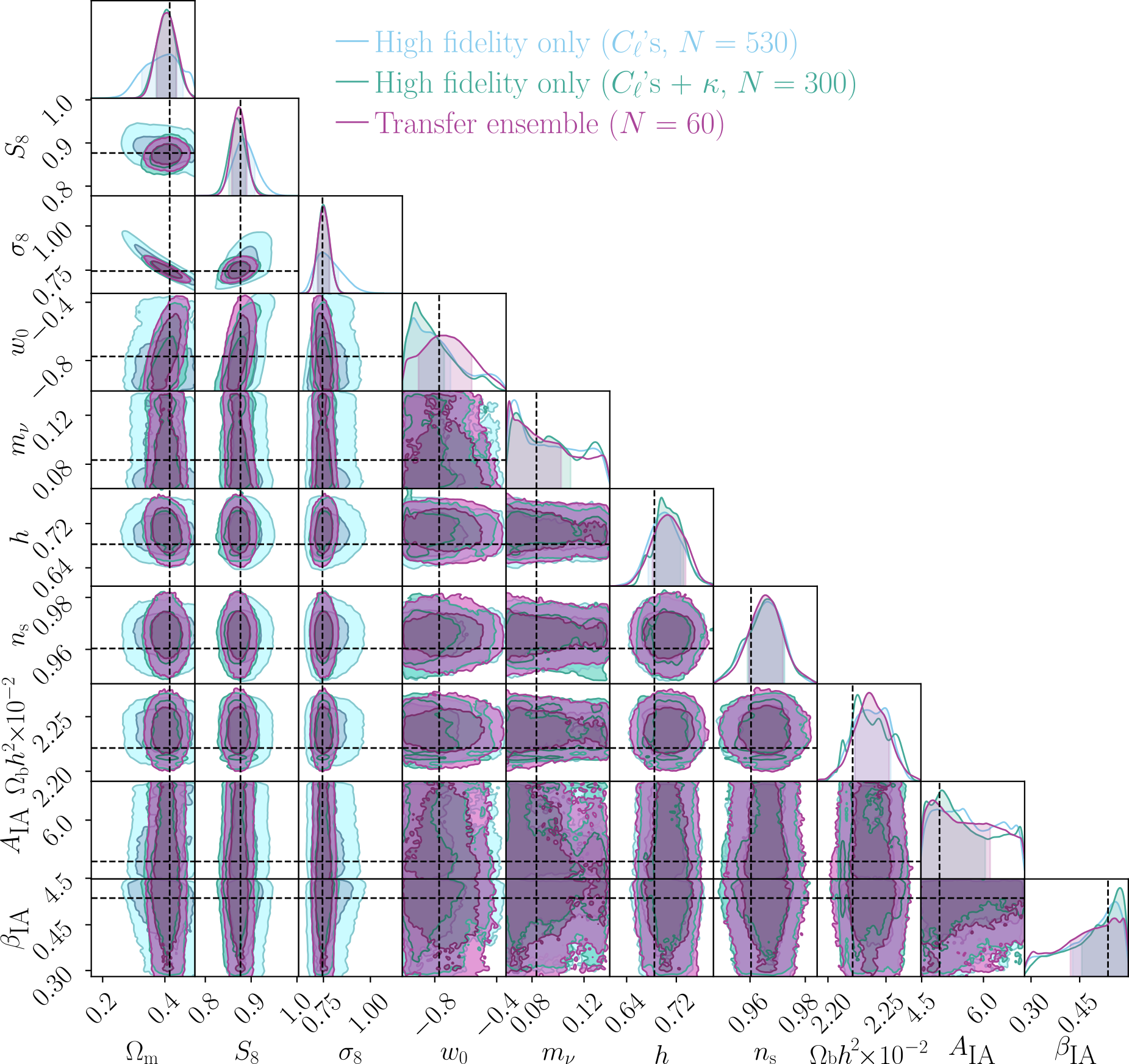}
    \caption{Inference example for the neural posterior estimation (NPE) approaches.  We present Gower Street only training (no pre-training), including and excluding higher-order information, against the results of our transfer learning ensemble. We include all 9 inference parameters as well as the derived parameter $S_8$. We have  selected an example where $\beta_\mathrm{IA}$ can be somewhat constrained beyond the prior by each model --- this is not always the case.}
    \label{fig:9param_posterior_example}
\end{figure*}

\noindent
\cref{fig:9param_posterior_example} shows an example of the full 9-D posteriors produced by representative models from each of the NPE approaches. 

\begin{figure}
\includegraphics[width=\columnwidth]{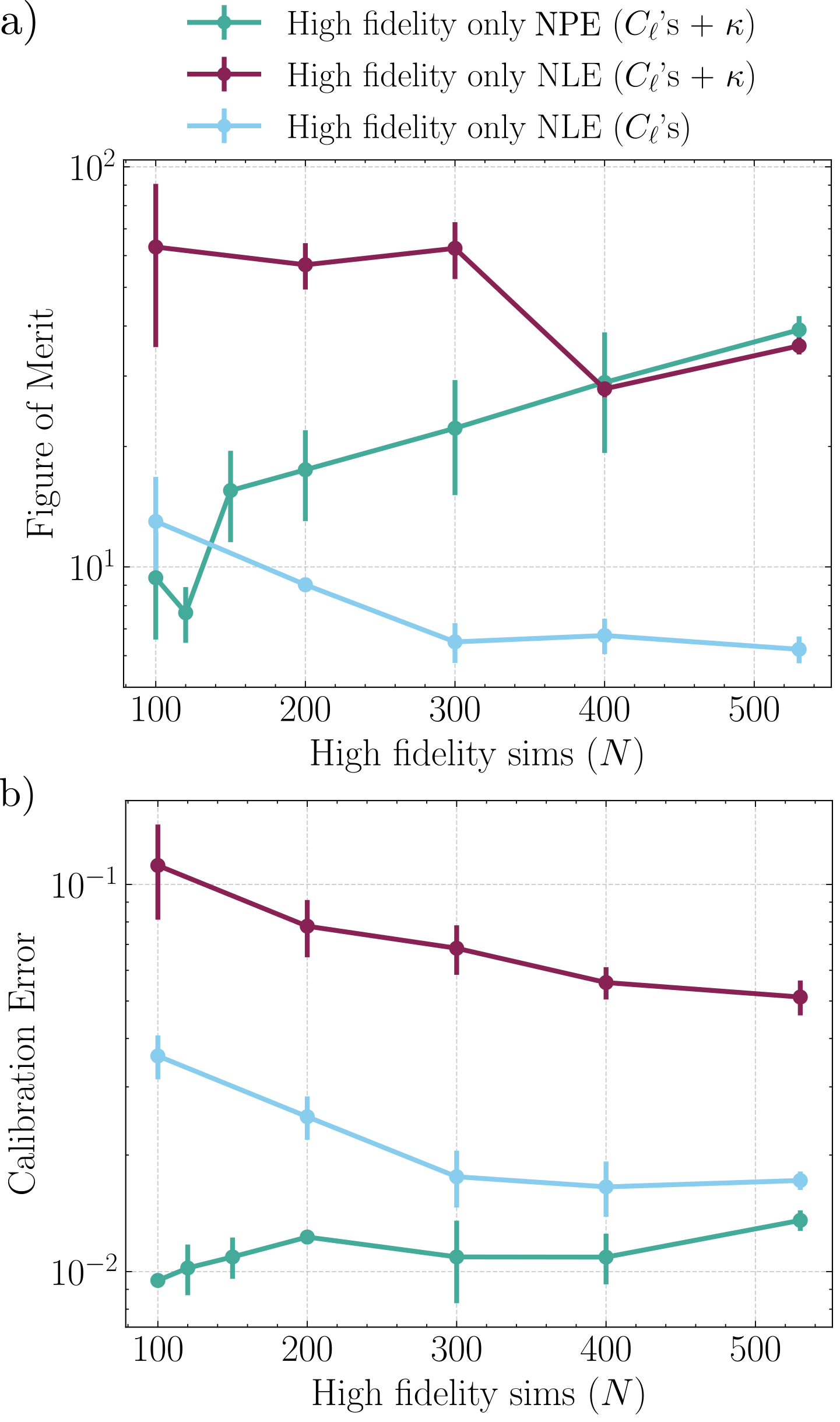}
    \caption{Inference results for the high-fidelity-only NLE approaches. We compare the NLE results from the beyond-two-point models ($\dim (t)=16$, maroon) against the two-point only models from the first stage of hybrid learning training (with $\dim (t)=8$, blue).}
    \label{app:NLE_bandpowers_metrics}
\end{figure}

\noindent
\cref{app:NLE_bandpowers_metrics} demonstrates the performance of the high-fidelity-only NLE models with a smaller summary statistic dimension, which mitigates but does not solve the calibration issues. 

\begin{figure*}
    \includegraphics[width=\textwidth]{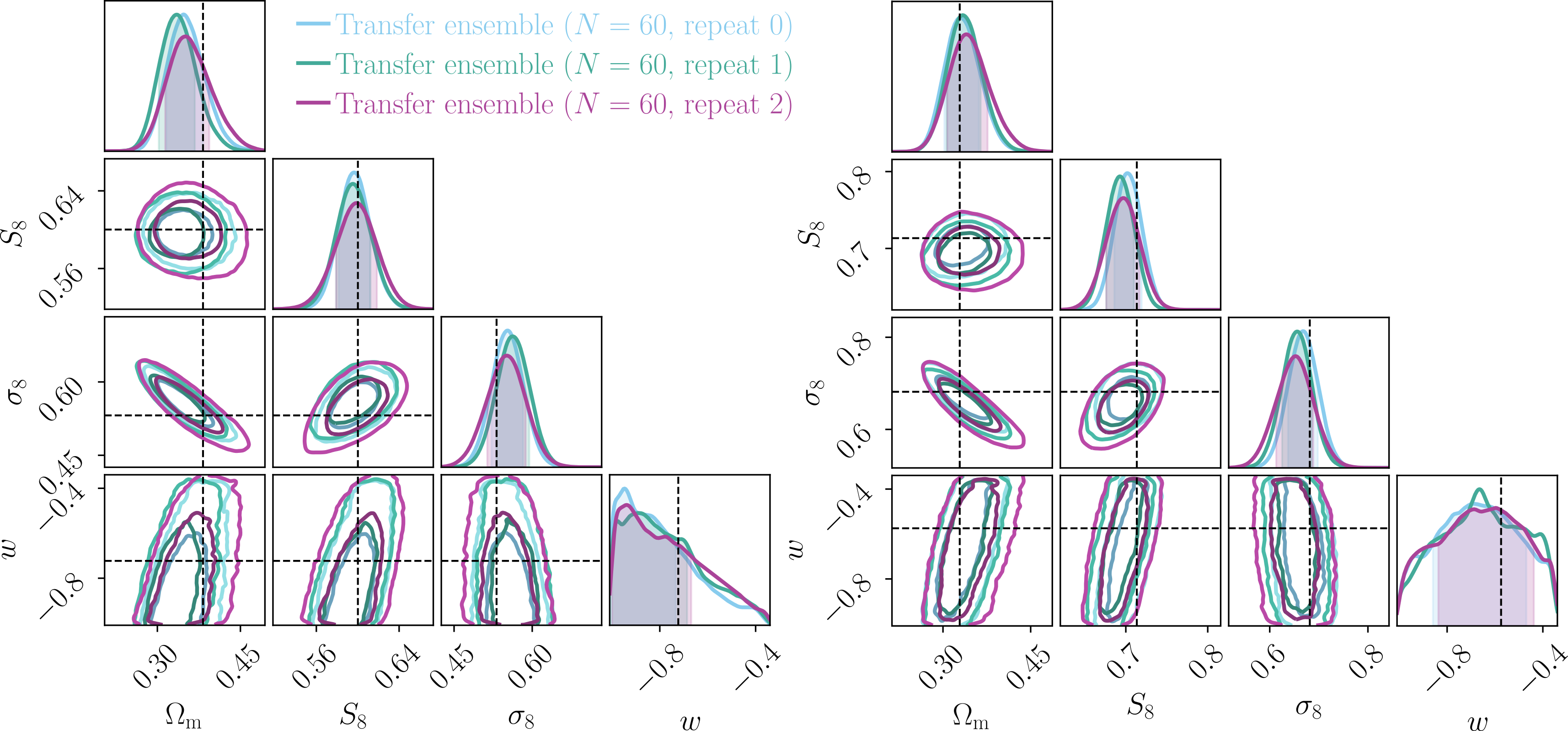}
    \caption{Two inference examples comparing the results from each of the three repeats of the NPE ensemble approach. Each ensemble was fine-tuned from independent \texttt{GLASS} pre-training runs. We expect minor differences due to the non-determinism of training and suboptimality of each learned neural compression model. For visualisation purposes we show the $\{\sigma_8, \Omega_\mathrm{m}, w\}$ subset of the modelled 9 parameter posterior distributions, along with the derived parameter $S_8$.  }
    \label{fig:repeated_training_example}
\end{figure*}

\noindent
\cref{fig:repeated_training_example} shows the intra-repeat variability of the NPE ensemble approach, which shows that independent training runs of the neural compressor and density estimation models lead to minor variability in posterior estimates.

\end{document}